\let\oldvec\vec% Store \vec in \oldvec
\documentclass[twocolumn]{aastex62}
\let\vec\oldvec% Restore \vec from \oldvec

\usepackage{graphicx}
\usepackage{amssymb,amsmath}
\usepackage{txfonts}
\usepackage{natbib}
\usepackage{rotating}
\usepackage{color, colortbl}
\usepackage{physics}
\usepackage[section]{placeins}
\renewcommand{\vec}[1]{\mathbf{#1}}

\definecolor{LightCyan}{rgb}{0.88,1,1}

\definecolor{Gray}{gray}{0.9}

\DeclareSymbolFont{mymathvariables}{OT1}{cmr}{m}{n}
\SetSymbolFont{mymathvariables}{normal}{OT1}{cmr}{m}{n}
\DeclareSymbolFontAlphabet{\mathnormal}{mymathvariables}
\DeclareMathSymbol{v}{\mathalpha}{mymathvariables}{118}

\newlength{\wfig}
\submitjournal{ApJ}
\shorttitle{Solar electron beam dynamics}
\shortauthors{Reid \& Kontar}

\begin{document}

\title{Spatial expansion and speeds of type III electron beam sources in the solar corona}

\correspondingauthor{Hamish Reid}
\email{hamish.reid@glasgow.ac.uk}

\author{Hamish A. S. Reid}
\affil{SUPA School of Physics and Astronomy \\
University of Glasgow, G12 8QQ, United Kingdom}

\author{Eduard P. Kontar}
\affil{SUPA School of Physics and Astronomy \\
University of Glasgow, G12 8QQ, United Kingdom}

   % \institute{SUPA School of Physics and Astronomy,
   % University of Glasgow, G12 8QQ, United Kingdom}

   % \date{}

   % \institute{}

\begin{abstract}

A component of space weather, electron beams are routinely accelerated in the solar atmosphere and propagate through interplanetary space.  Electron beams interact with Langmuir waves resulting in type III radio bursts. Electron beams expand along the trajectory, and using kinetic simulations, we explore the expansion as the electrons propagate away from the Sun. Specifically, we investigate the front, peak and back of the electron beam in space from derived radio brightness temperatures of fundamental type III emission.  The front of the electron beams travelled at speeds from 0.2c--0.7c, significantly faster than the back of the beam that travelled between 0.12c--0.35c.  The difference in speed between the front and the back elongates the electron beams in time.  The rate of beam elongation has a 0.98 correlation coefficient with the peak velocity; in-line with predictions from type III observations.  The inferred speeds of electron beams initially increase close to the acceleration region and then decrease through the solar corona.  Larger starting densities and harder initial spectral indices result in longer and faster type III sources.   Faster electron beams have higher beam energy densities, produce type IIIs with higher peak brightness temperatures and shorter FWHM durations.  Higher background plasma temperatures also increase speeds, particularly at the back of the beam.  We show how our predictions of electron beam evolution influences type III bandwidth and drift-rates.  Our radial predictions of electron beam speed and expansion can be tested by the upcoming in situ electron beam measurements made by Solar Orbiter and Parker Solar Probe.
\end{abstract}


% \maketitle

% \pagebreak

\section{Introduction}

Solar electron beams, accelerated via magnetic instabilities in the solar atmosphere, do not simply propagate scatter-free through the solar corona and interplanetary space. Beams interact with the background plasma which results in them not propagating at a constant velocity. One such resonant interaction is the excitation of Langmuir waves through the bump-in-tail instability \citep{Ginzburg:1958aa}.  Electron beams and Langmuir waves are detected in situ together \citep{Gurnett:1975aa,Gurnett:1976aa}, along with type III radio bursts that are generated via wave-wave interactions by the Langmuir waves. Remote sensing type III bursts at high frequencies provides information about electron beam speeds in the solar atmosphere where in situ measurements are not currently possible. The energy exchange between the electron beam and Langmuir waves modifies the electron distribution
\citep[e.g.][]{Zaitsev:1972aa,Magelssen:1977aa,Melnik:1998aa,Kontar:2001ad}.

% Type III radio bursts are a consequence of the excited Langmuir waves produced by propagating electron beams.  The Langmuir waves emit radio waves near the plasma frequency and these radio waves can be remotely detected at, and near the Earth.  Because the plasma frequency decreases as a function of distance from the Sun, the change in type III frequency with time (drift rate) provides information about how fast the electron beam travels through the heliosphere.  The frequency width of type IIIs are directly related to the distribution of Langmuir waves in space, and hence the distribution of the electron beam in space.  If we understand the physics that dictates electron beam speed and length then we can use type III bursts as a remote sensor of electron beam properties, both initial and during transport, and the properties of the background coronal and heliospheric plasma.

To estimate electron beam velocities via remote sensing, the frequency drift rate $\pdv{f}{t}$ of type III bursts is often used, where $f$ is the frequency. Because the plasma frequency decreases as a function of distance from the Sun, the frequency drift rate provides information about how fast the electron beam travels through the heliosphere. Type III drift rates decrease in magnitude as a function of decreasing frequency. This is typically related to the decreasing magnitude of the background electron density gradient at farther distances from the Sun \citep{Kontar:2001ab,Ratcliffe:2012aa}.  A power-law dependence over four orders of magnitude was found by \citet{Alvarez:1973ab} combining results from numerous studies. The drift rates have been reported to be slightly lower at higher frequencies \citep[e.g.][]{Achong:1975aa,Melnik:2011aa}.  Drift rates are usually found using the evolution of the peak flux in time although the onset time is occasionally used.  Type III drift rates using the onset time have been shown to be faster than when the peak time is used, which in turn are faster than when the decay time is used \citep{Reid:2018aa}.  Type III drift rates are also influenced by the radio emission mechanism that converts Langmuir waves to radio waves and the subsequent radio wave propagation from source to observer \citep[e.g.][]{Kontar:2017ab}.  However, in this work we concentrate on how the beam dynamics influences the type III drift rate.

Velocities of electron beams deduced from type III bursts are measured in fractions of the speed of light. Typical velocities are between 0.2--0.5~c \citep[e.g.][]{Wild:1959aa,Alvarez:1973ab,Suzuki:1985aa,Reid:2014ab} although relativistic velocities ($>0.6$~c) have been found in the corona \citep{Poquerusse:1994aa,Klassen:2003aa}. 
A recent study \citet{Reid:2018aa} of 31 type III bursts estimated velocities deduced from the rise, peak and decay times of type III bursts to be 0.2c, 0.17c and 0.15c respectively.  Exciter speeds were found to be significantly lower in the interplanetary medium from type III bursts below 20~MHz \citep{Fainberg:1972aa,Dulk:1987aa,Krupar:2015aa,Reiner:2015aa}, going down to 0.1~c or below.   Electron beams were found to decelerate on their way to the Sun; a constant velocity was not a good approximation. Typical deceleration values below 20~MHz have been found around 10~km~s$^{-2}$ \citep{Krupar:2015aa}.    

Whilst a single exciter velocity is typically attributed to electron beams from type III bursts, the Langmuir wave generating beams have a broad distribution in energy space \citep[e.g.][]{Zaitsev:1972aa}.  The electron beam and the resonating Langmuir waves can be described as a beam-plasma structure using gas-dynamic theory \citep[e.g.][]{Ryutov:1970aa,Melnik:1995aa,Kontar:1998aa,MelNik:1999ab,Kontar:2003aa}. The electron and Langmuir waves travel together through a constant exchange of energy; the electron distribution relaxing to a plateau in velocity space between $v_{\rm min}$ and $v_{\rm max}$.  It was shown that the resulting beam-plasma structure \citep{MelNik:1999ab,Kontar:2003aa} moves through space with the average velocity of the electrons $v_{\rm bps}=(v_{\rm max}+v_{\rm min})/2$,where $v_{\rm max}$ is the speed of electrons at the injection and $v_{\rm min}$ is the minimum electron speed.  Electron beams injected at the Sun are not necessarily Maxwellian, but the collective behaviour exhibited by the beam-plasma structure helps explain why near-constant beam velocities are inferred from type III bursts \citep[e.g.][]{MelNik:1999ab,Kontar:2003aa}.

Simulations of electron beams propagating through the solar corona and interplanetary space, interacting with Langmuir waves, have been carried out for decades \citep[e.g.][]{Takakura:1976aa,Magelssen:1977aa}.  For simulations over significant distances, the quasilinear description\citep[e.g.][]{Vedenov:1963aa} is typically used.  From observations of electrons in situ at 1~AU \citep{Lin:1981aa,Krucker:2007aa} and from estimations from X-ray observations \citep{Holman:2011aa}, a power-law electron beam is commonly assumed as the injection function.  Previous numerical studies \citep[e.g.][]{Reid:2013aa,Ratcliffe:2014aa,Li:2014aa} have shown that, over large distances of a few solar radii or more, the velocity range of electrons which generates the bulk of the Langmuir waves decreases.  This is consistent with the decrease in type III drift rate as a function of frequency \citep{Fainberg:1972aa,Krupar:2015aa}.

% Recent simulations have looked at beams propagated down through the solar atmosphere \citep[e.g.][]{Hannah:2009aa,Hannah:2013aa}, in regions of constant density \citep[e.g.][]{Khalilpour:2013aa,Khalilpour:2014aa,Khalilpour:2015aa} and up through the solar corona \citep[e.g.][]{Kontar:2001ab,Foroutan:2008aa,Kontar:2009aa,Khalilpour:2012aa,Reid:2013aa,Li:2013aa,Ratcliffe:2014aa,Reid:2015aa,Reid:2017ab}.  The properties of the electron beam at they propagate depend upon the power-law height (starting density) and slope (spectral index), with an increase in the phase space density of high velocity electrons typically increasing electron beam speeds.  As electrons travel out from the Sun, the characteristic velocity of the electron beam, as indicated from synthetic radio emission, has been found to decrease as a function of distance from the acceleration.  A combination of Langmuir wave refraction and Landau damping removes energy from the beam-plasma structure whilst the electron beam decreases in density as it follows an expanding magnetic field.

If electrons propagated freely, the beam length would increases through velocity dispersion as a beam consists of electrons with a range of velocities. As faster electrons outpace slower electrons, the fastest electrons will travel a distance of $v_{\rm max}\Delta t$ whilst the slowest electrons will travel a distance of $v_{\rm min}\Delta t$.  The electron beam length will increase over time by $(v_{\rm max}-v_{\rm min})\Delta t$ over some time $\Delta t$.  However, evidently, free electron propagation is an oversimplification, and electron velocities will change during propagation through wave-particle interactions and/or pitch-angle scattering.  How electron beam length thus develops as a function of time is poorly understood.

The type III parameter most associated with the electron beam length is the instantaneous bandwidth; the width in frequency between the minimum and maximum frequency at any one point in time.  The type III bandwidth provides information about where the electron 
beam is able to radiate at any given time.  The type III bandwidth has not been extensively studied in the past.  \citet{Hughes:1963aa} defined the bandwidth from the leading edge to the highest frequency, finding a bandwidth of 100~MHz from a leading edge of 100~MHz.  \citet{Melnik:2011aa} defined the instantaneous bandwidth as the half-power bandwidth, finding a bandwidth of 15~MHz around 18~MHz for powerful type III bursts. To account for asymmetric bandwidth profiles, \citet{Reid:2018aa} defined the bandwidth using the rise and decay times of the type III emission to estimate $f_{\rm min}$ and $f_{\rm max}$ for each point in time.  The instantaneous bandwidth $\Delta f = f_{\rm max} - f_{\rm min}$ at the frequency with roughly peak intensity.  Bandwidths of 20~MHz were found around 50~MHz.  Both \citet{Melnik:2011aa} and \citet{Reid:2018aa} found the bandwidth to vary roughly as $\Delta f=0.6f$.

In this study we investigate both the velocity of electron beams and their length as the beam travels through the solar corona.  We start by defining different regions of the electron beam, the front, peak and back of the beam.  We then show how these parameters evolve with time and what causes their evolution in Section \ref{sec:dynamics}.  How the change in the initial beam parameters affects the beam velocity and length are shown in Section \ref{sec:params}, along with the effect of the background plasma electron temperature.  We explore the effects on type III properties from injecting electron beams with different parameters in Section \ref{sec:band_drift}.  The results are discussed in Section \ref{sec:discussion}, with conclusions provided in Section \ref{sec:conclusions}.

\section{Simulation model} \label{sec:model}

\subsection{Kinetic model}

To investigate the length and velocity of an electron beam we used 1D self-consistent kinetic model \citep{Kontar:2001aa} of the electron distribution function $f(v,r,t)$ [electrons $\rm{cm}^{-4}~s$] and the spectral energy of Langmuir waves $W(v,r,t)$ [ergs $\rm{cm}^{-2}$].  The model calculated the 1D propagation of the electrons along the direction of the guiding magnetic field in the WKB approximation, where waves are treated as quasi-particles interacting resonantly ($\omega_{pe}=kv$, where $\omega_{pe}$ is the plasma frequency) with the electrons.  

An in-depth description of the kinetic equations, source model and background plasma model can be found in \citet{Reid:2017ab}.  The same background plasma density model, obtained using the equations for a stationary spherical symmetric solution \citep{Parker:1958aa} with a normalisation factor by \citet{Mann:1999aa}, was used in all simulations.  No density fluctuations were added.  The source function parameters do vary between simulations.  The source function is given by 
\begin{equation}\label{eqn:source}
S(v,r,t) = A_v v^{-\alpha}\exp\left(-\frac{r^2}{d^2}\right)A_t\exp\left(-\frac{(t-t_{inj})^2}{\tau^2}\right),
\end{equation}
where $A_v=n_b(\alpha-1)/(v_{\rm min}^{1-\alpha}-v_{\rm max}^{1-\alpha})$ is a normalisation constant that sets the beam density $n_b$ [cm$^{-3}$].  The constant $A_t=(\tau\sqrt{\pi})^{-1}$ normalises the integral of the exponential involving time to one. The characteristic variables that define the behaviour of the source function in velocity, distance and time are $\alpha$, $d$ [cm] and $\tau$ [s], respectively.

\subsection{Radio emission} \label{sec:radio}

We approximate a dynamic spectrum of fundamental emission from the Langmuir wave spectral energy density assuming a saturation level of plasma emission \citep{Melrose:1980aa,Tsytovich:1995aa,Lyubchyk:2017aa}. 
Beam-driven Langmuir wave growth causes $W_L >> W_S$, 
where $W_L$, $W_S$ are the spectral energy density of Langmuir waves 
and ion-sound waves, respectively. 
The nonlinear decay $L\rightarrow T+S$ causes exponential growth of ion-sound waves \citep[e.g.][]{Melrose:1980aa}.  This will increase both $W_S$ and the brightness temperature $T_T$ until such time as $W_S >> W_L$.  At this point the process saturates at the level
\begin{equation} \label{eqn:t_b_orig}
T_T(k,r,t) = \omega_T(r) \frac{\eta}{k_s(r)^2 k_b}\frac{W_L(k,r,t)}{\omega_L(r)}.
\end{equation}
where $k_s$ is the wavenumber of the ion sound waves, $\omega_L, \omega_T$ are the Langmuir wave and electromagnetic wave angular frequencies and $k_b$ is the Boltzmann constant. If we assume that $\omega_T\approx \omega_L$ and $k_s\approx k_L$ then for $\eta=(2\pi)^2$ we obtain 
\begin{equation} \label{eqn:t_b}
k_b T_T(k,r,t) \approx \frac{(2\pi)^2}{k_L(r)^2}W_L(k,r,t).
\end{equation}
To obtain the brightness temperature at each position (frequency), $T_T(r,t)$, we use the peak value of $T_T(k,r,t)$ for each point in space and time. The spread in $k$ (and consequently frequency) is small for each point in space and the peak value gives very similar results to using the mean value of $T_T(k,r,t)$ as a function of $k$.

\begin{figure}\center
\includegraphics[width=\wfig,trim=30 0 0 18,clip]{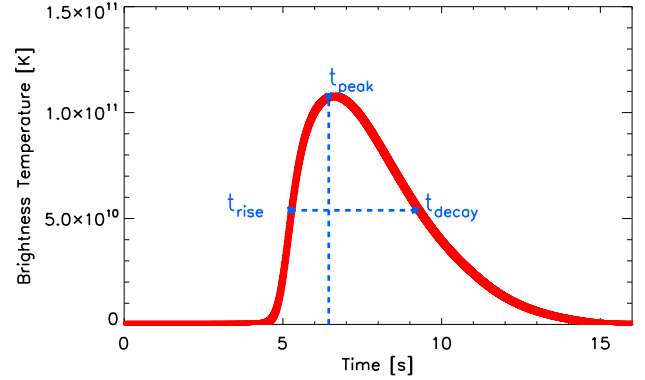}
\caption{Time profile of the radio brightness temperature deduced from the Langmuir waves at 30~MHz or $0.76~R_\odot$.  The rise, peak and decay times corresponding to the peak value and half-width half-maximums are indicated.}
\label{figure:time_profile} 
\end{figure}

\section{Electron beam dynamics} \label{sec:dynamics}

\vspace{20pt}
\begin{center}
\begin{table*}
\centering
\caption{Initial beam parameters for the electron beam injected into the solar corona.}
\begin{tabular}{ c  c  c  c  c  c  c }

\hline\hline

Energy Range & Velocity Range & Spectral Index & Injection Time & Beam Size & Source Height & Density Ratio \\ \hline
$0.28-125$~keV &  $2.6-54~v_{\rm th}$ & $\alpha=8.0$ & $\tau=0.001$~s & d=$10^9$~cm & h=$3\times 10^9$~cm &  $n_b/n_e=0.003$  \\

\hline
\end{tabular}
\vspace{20pt}
\label{tab:beam_sun}
\end{table*}
\end{center}

When diagnosing electron beam dynamics from type III radio bursts, we are getting information about electrons that are undergoing significant wave-particle interactions with the Langmuir waves, which subsequently drive the radio emission.  The bulk of the energy in the system is contained within the electrons and the Langmuir waves; the radio emission is not energetically important.  Therefore, the motion of the electron beam through the heliosphere can be modelled using the electrons and Langmuir waves alone.

\begin{figure*}\center
\includegraphics[width=0.90\textwidth]{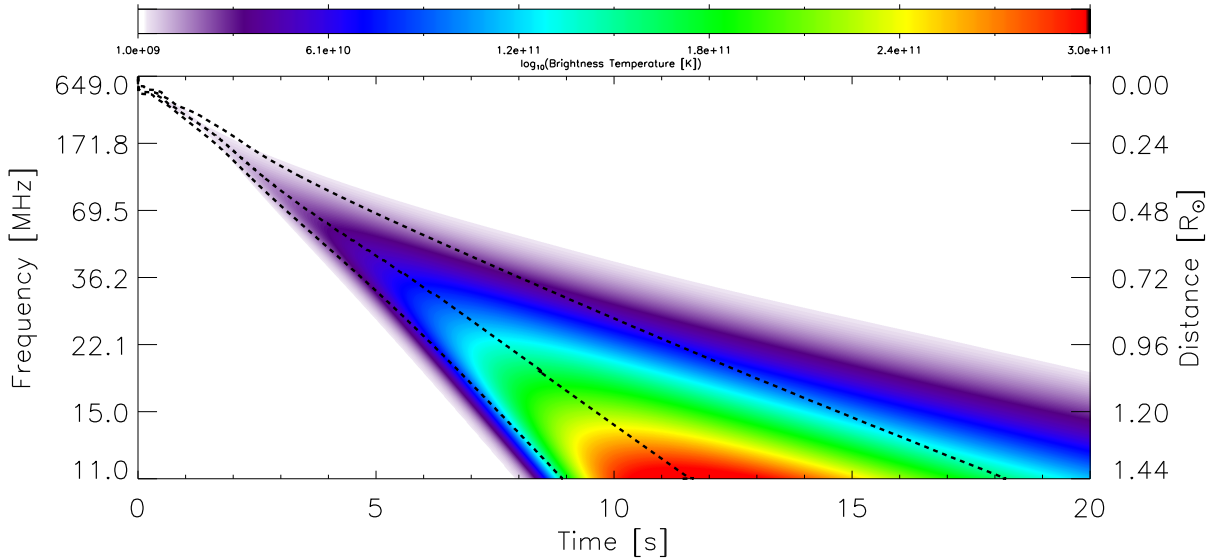}
\caption{Type III brightness temperature dynamic spectrum assuming fundamental emission and a saturation of ion sound waves, found using Equation \ref{eqn:t_b} and an initial electron beam described in Table \ref{tab:beam_sun}.   The rise, peak and decay times from the FWHM at each frequency are indicated by black dashed lines.}
\label{figure:dynspec} 
\end{figure*}

In this work, we investigate the electron beam using the time profile of the fundamental radio emission that is estimated from the Langmuir waves (Section \ref{sec:radio}).  For each position (frequency) we find the time, $t_{\rm peak}$, at which the radio brightness temperature is highest.  We then find the times which correspond to the half-width half-maximum of the intensity profile, $t_{\rm rise}$ and $t_{\rm decay}$, of $T_T(r,t)$.  Figure \ref{figure:time_profile} shows a sample time profile of $T_T(r,t)$ at 30~MHz ($0.76~R_\odot$) using the simulation parameters given in Table \ref{tab:beam_sun}, illustrating $t_{\rm rise}$, $t_{\rm peak}$ and $t_{\rm decay}$.  For each time $t$, the peak of the electron beam is defined by the position where $t\approx t_{\rm peak}$.  Similarly, the front and back of the electron beam are at positions where $t\approx t_{\rm rise}$ and $t\approx t_{\rm decay}$, respectively.

\begin{figure*}\center
\includegraphics[width=0.90\textwidth]{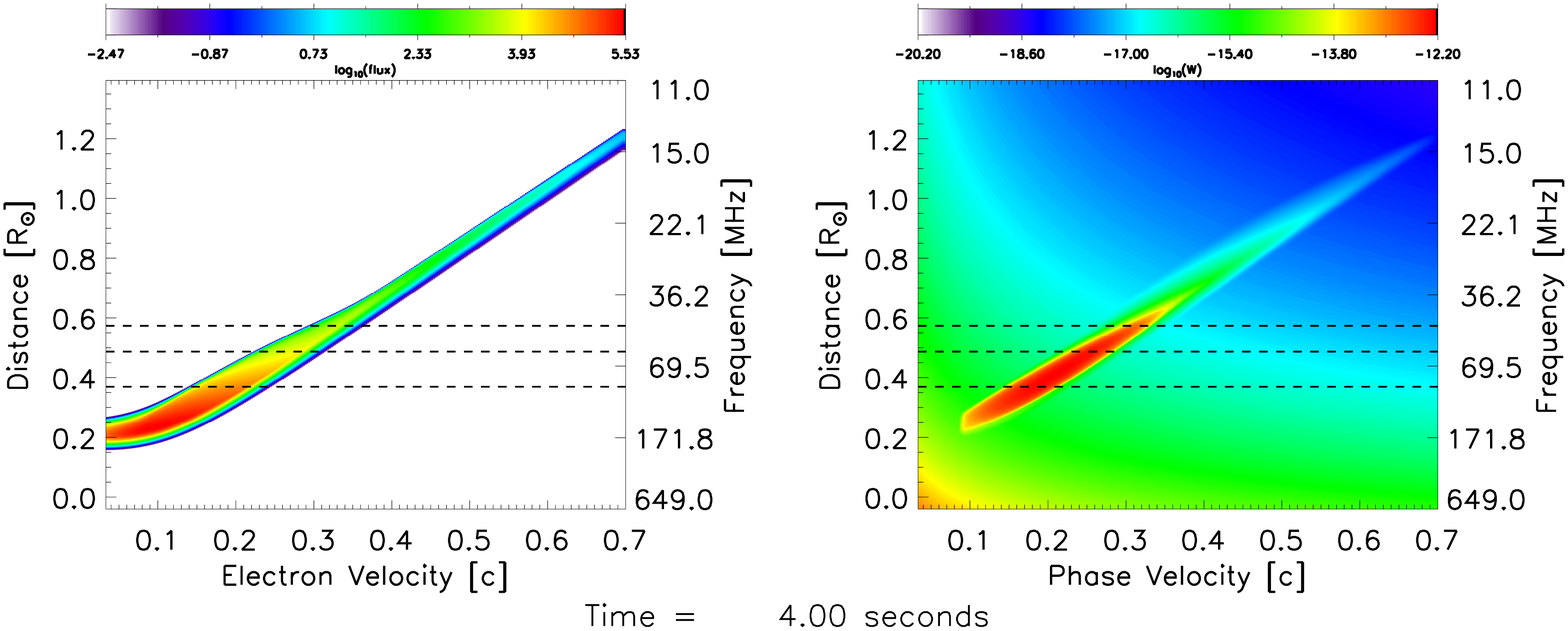}
\includegraphics[width=0.90\textwidth]{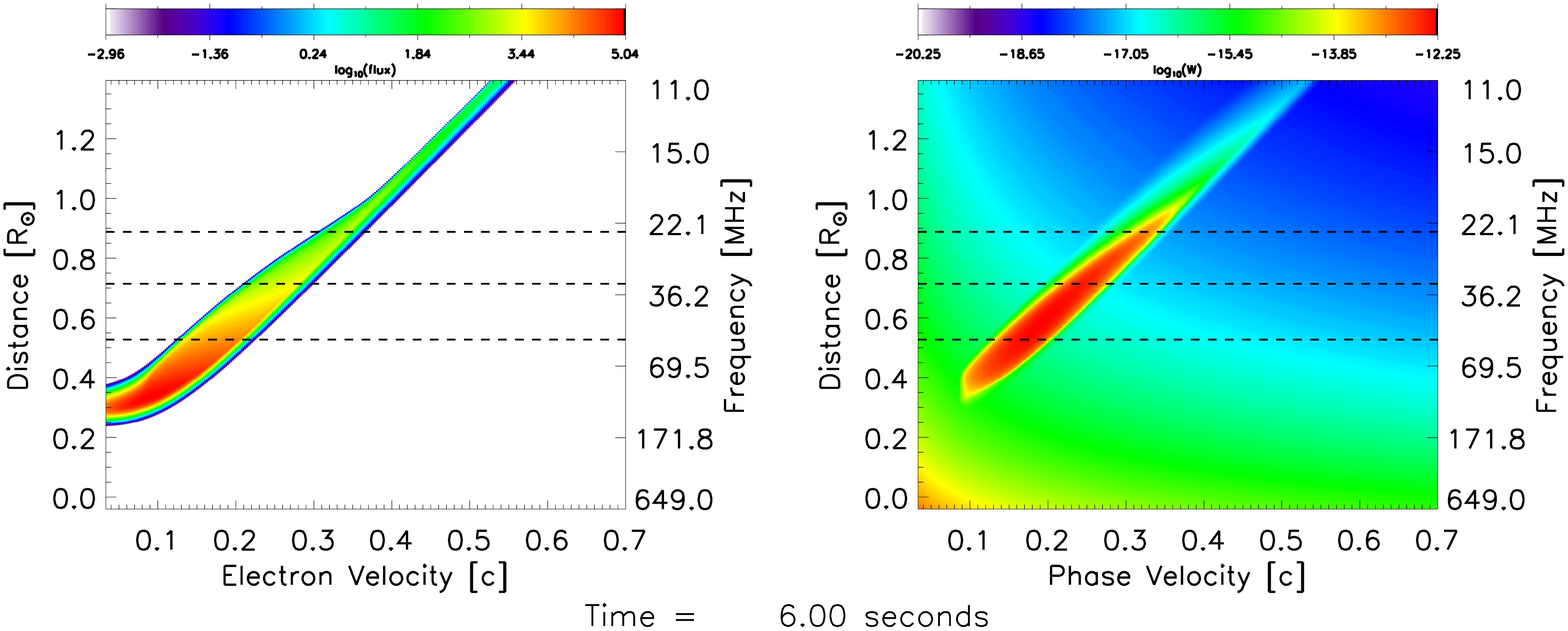}
\includegraphics[width=0.90\textwidth]{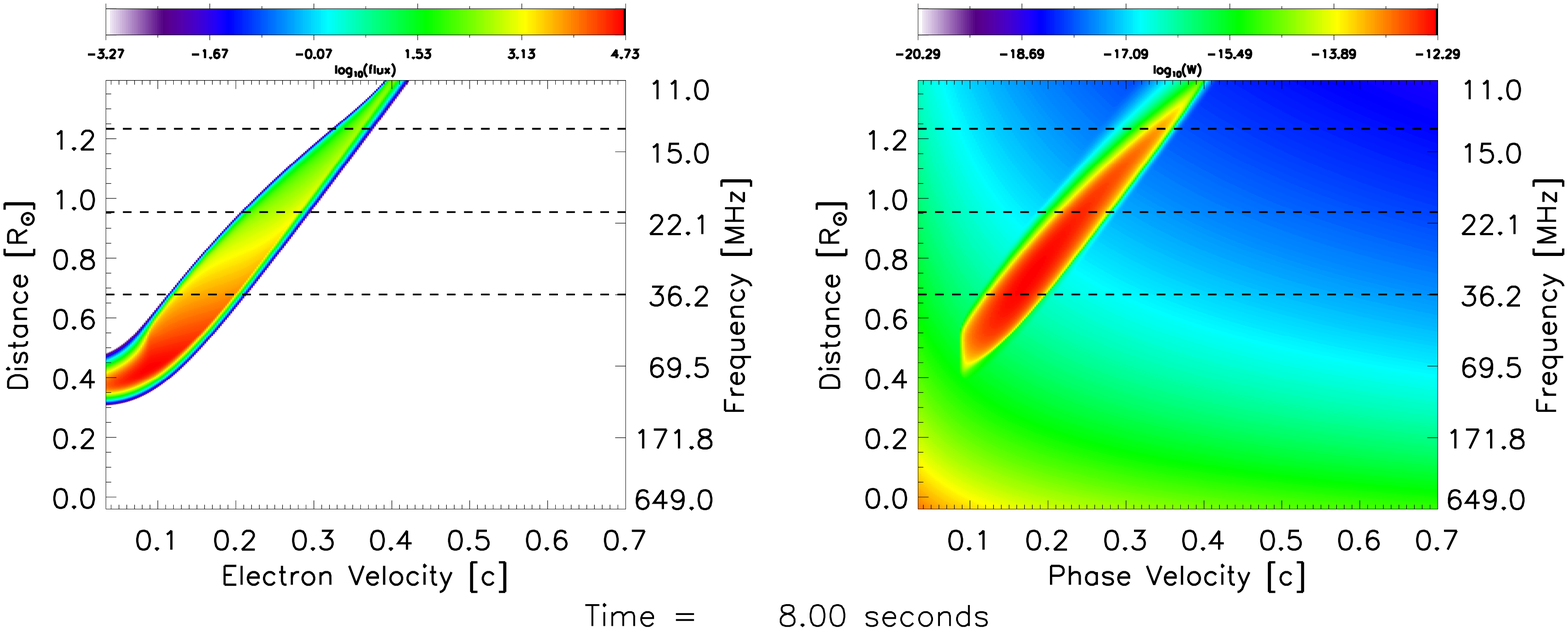}
\caption{The electron flux and Langmuir wave spectral energy density at 4, 6, 8 seconds after injection for the initial electron beam parameters given in Table \ref{tab:beam_sun}.  The corresponding front, peak and back of the electron beam is shown by horizontal dashed lines, found from the derived radio brightness temperature of fundamental type III emission.}
\label{figure:eflux} 
\end{figure*}

\subsection{Electron beam properties}

We initially inject an electron beam into the solar corona using the source function given in Equation \ref{eqn:source} with parameters given in Table \ref{tab:beam_sun}.  The background temperature is set as 1~MK, giving a thermal velocity of $v_{\rm th}=\sqrt{k_bT_e/m_e}=3.9\times10^8~/rm{cm~s}^{-1}$.  The minimum energy (velocity) is sufficiently close to the thermal distribution that Langmuir waves are significantly damped.  The maximum energy (velocity) is low enough to ignore relativistic effects (the Lorentz factor is at most 1.4), but high enough to cover the bulk of relevant energies for electrons that induce type III bursts.  The spectral index is typical for accelerated electron beams, found from X-ray observations \citep{Holman:2011aa}.  The size and height are within the range of estimates found from X-ray and radio flare observations  \citep{Reid:2014aa}.  The density ratio is given at the source height and corresponds to an initial background electron density of $n_e=3\times10^{9}~\rm{cm}^{-3}$ (500~MHz) and an initial beam density of $n_b=10^7~\rm{cm}^{-3}$.

Figure \ref{figure:dynspec} shows the dynamic spectrum of fundamental emission produced by the beam-driven Langmuir waves.  The general trend of increasing intensity with decreasing frequency is shown \citep[e.g.][]{Dulk:1998aa,Krupar:2015aa,Reid:2017aa}.  The rise, peak and decay times, found from the full-width half-maximum (FWHM) of the brightness temperature at each frequency are indicated.

To demonstrate how the rise, peak and decay times of the radio emission correspond to the front, peak and back of the electron beam, phase space snapshots of the electron distribution function, in units of electron flux [$\rm{cm}^{-2}~\rm{s}^{-1}~\rm{eV}^{-2}$], and the Langmuir wave spectral energy density [$\rm{ergs}~\rm{cm}^{-2}$] are shown in Figure \ref{figure:eflux}.  We have indicated the front, peak and back of the beam as horizontal dashed lines.

The radio emission indicates the bulk of the electron beam in phase space, except from two parts.  The first part is the very front of the electron beam.  The small electron flux causes the growth rate of Langmuir waves to be insignificant compared to the background level.  These electrons consequently do not diffuse in velocity space and undergo ballistic transport in our model.  The second part is the back of the beam.  The electron velocities are so low that Langmuir waves are strongly Landau-damped by the background plasma.

\begin{figure}\center
\includegraphics[width=\wfig,trim=40 0 20 18,clip]{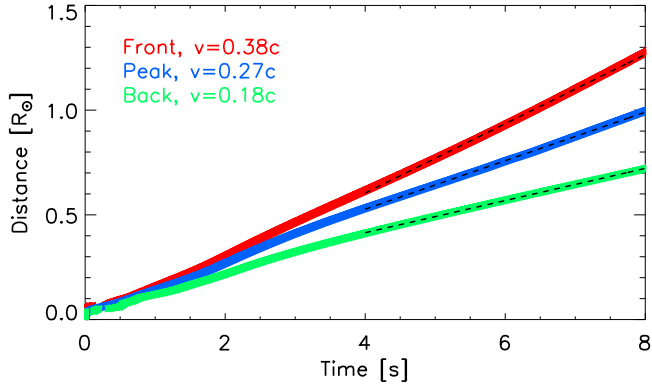}
\caption{Position of the front, peak and back of the electron beam as a function of time.  The dashed lines are linear fits to these positions from 3.5 to 8 seconds, with the gradient providing a constant velocity, indicated in the plot legend.}
\label{figure:sim_FWHM}
\end{figure}

The evolution of the front, peak and back of the electron beam, found from radio emission, are presented in Figure \ref{figure:sim_FWHM} as a function of time.  We can apply a linear fit to the positions as a function of time after $3.5$ seconds to approximate the velocities of the front, peak and back of the electron beam.  We find that the front travels faster than the peak that in turn travels faster than the back, as one might expect.  The fits are shown in Figure \ref{figure:sim_FWHM} along with the derived velocities.   

\begin{figure}\center
\includegraphics[width=\wfig,trim=40 0 20 18,clip]{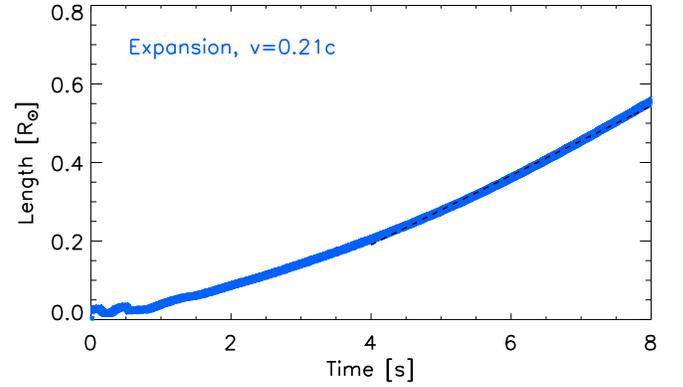}
\caption{Length of the electron beam, found from the distance between the front and back of the electron beam.  The dashed line is a linear fit to the length from 4 to 8 seconds, with the gradient providing a constant velocity for the beam expansion, indicated in the plot legend.}
\label{figure:length}
\end{figure}

We define the length of an electron beam as the distance between the front and back of the electron beam.  The electron beam length increases as a function of time, shown in Figure \ref{figure:length}.  A linear fit to the length as a function of time gives a constant expansion velocity (increase in length as a function of time).  The expansion velocity, $\Delta v$, is essentially the difference between the velocity at the front of the beam $v_f$ and the velocity at the back of the beam $v_b$ such that $\Delta v\approx v_f-v_b$.

\subsection{Electron beam velocity evolution} \label{sec:vel_evolve}

\begin{figure}\center
\includegraphics[width=\wfig,trim=40 0 20 18,clip]{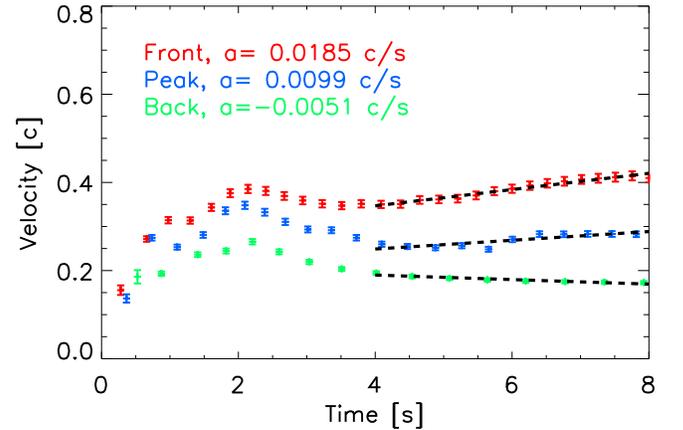}
\caption{Velocity of the front, peak and back of the electron beam, found from the derivative of Figure \ref{figure:sim_FWHM}.  The dashed line is a linear fit to the velocities from 4 to 8 seconds, with the gradient providing a constant acceleration, indicated in the plot legend.}
\label{figure:vel_time}
\end{figure}

The velocity of the front, peak and back of the electron beam is not actually constant but changes as a function of time, captured in Figure \ref{figure:vel_time}.  To approximate the velocity from Figure \ref{figure:sim_FWHM} we used the gradient of a straight line fit over 30 points in space, pertaining to a distance of $3\times10^9$~cm, with the fitting errors shown.  After around four seconds, the velocity of the front and the peak of the beam increases as a function of time.  This is related to a gradual increase in the velocity of the electrons that resonate with Langmuir waves; caused by physical processes including density inhomogeneity, radial expansion and the decreasing background electron density gradient.  The increase in velocity does not continue throughout the solar wind and will begin to decrease after the electron beam becomes more dilute at farther distances from the Sun.  The velocity of the back of the beam decreases as a function of time.  Conversely, this is related to a gradual decrease in the velocity of electrons that resonate with the Langmuir waves.  

\begin{figure}\center
\includegraphics[width=\wfig,trim=40 0 0 18,clip]{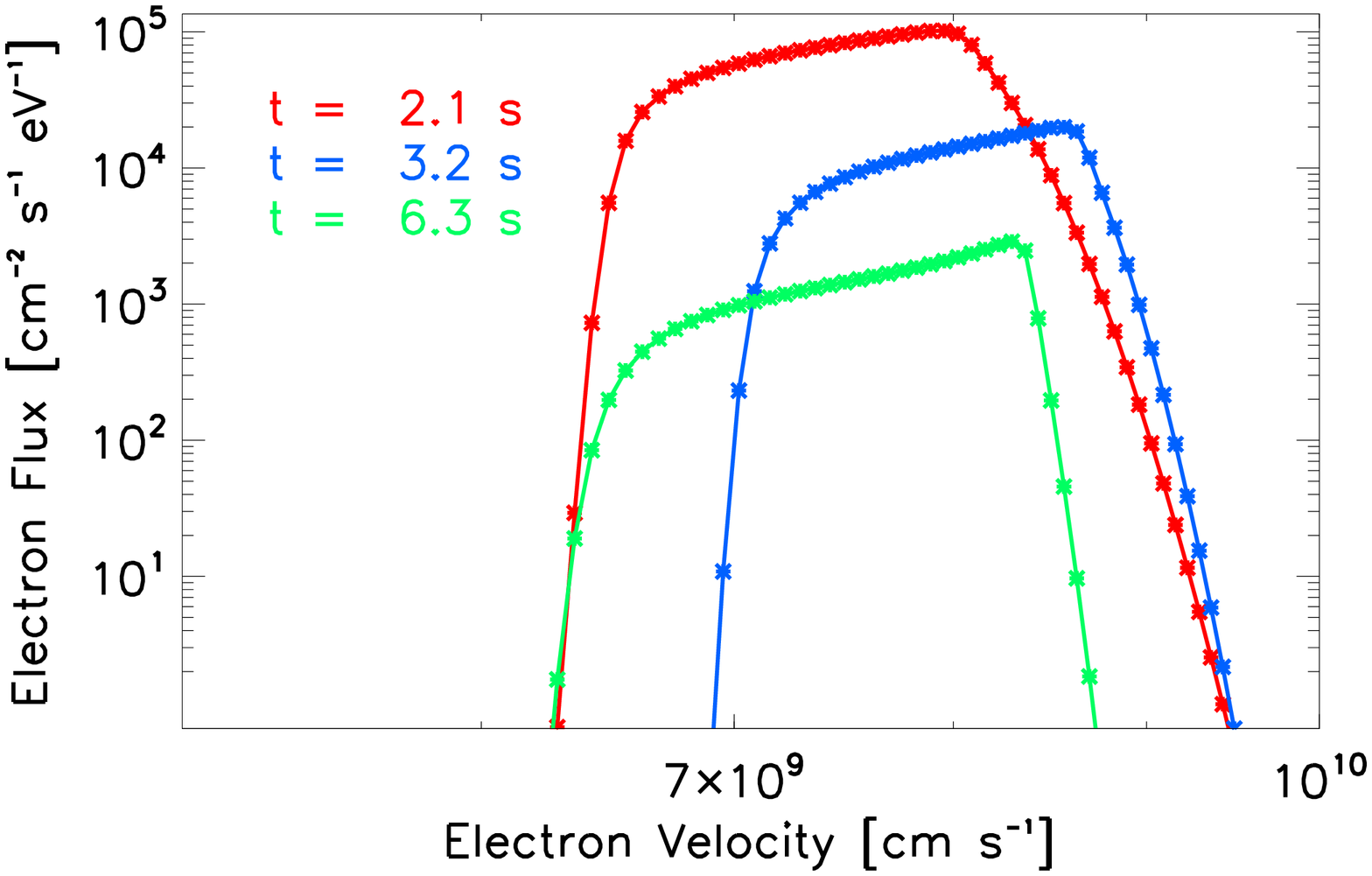}
\includegraphics[width=\wfig,trim=15 0 0 18,clip]{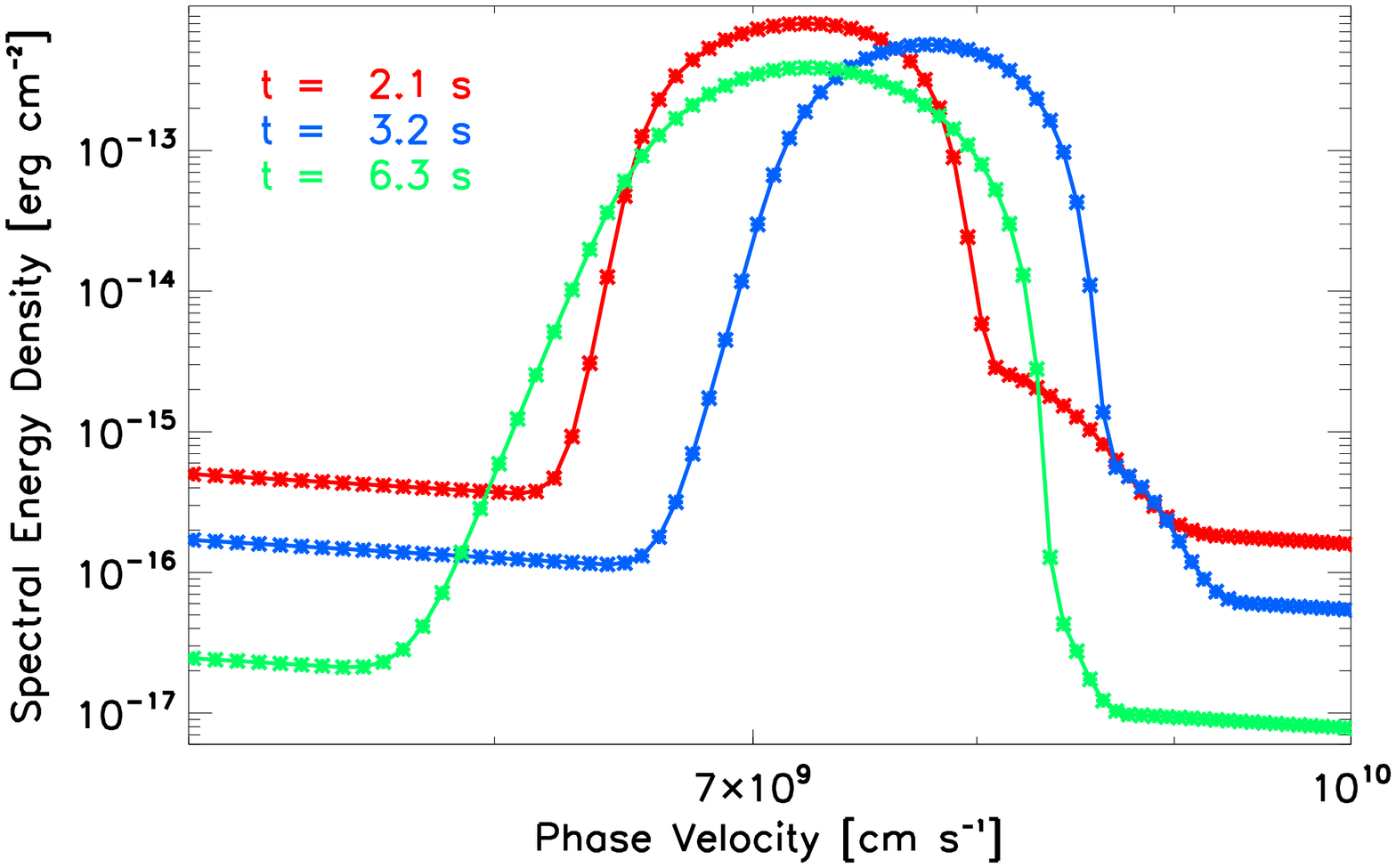}
\caption{Electron flux and Langmuir wave spectral energy density as a function of velocity.  The different times show the increase and subsequent decrease in wave-particle resonant velocities.}
\label{figure:f_vel_time}
\end{figure}

The velocity profile before four seconds is more complex, showing a sharper increase and subsequent decrease in velocity.  At the earliest times the electrons that generate a substantial amount of Langmuir waves are at lower velocities.  As $t$ increases, these electrons are at higher velocities until some time $t_p$.  At times $t>t_p$, the velocities of the electrons that produce substantial Langmuir waves become lower.  The change in the relevant velocities of the electrons that produce the highest Langmuir wave spectral energy density is demonstrated in Figure \ref{figure:f_vel_time}.  The electron flux and Langmuir wave spectral energy density as a function of velocity at different times are shown, where $t_p\approx 3.2$~s.  The positions, $r$, of $f(v,r,t)$ and $W(v,r,t)$ correspond to where the spectral energy density is at a maximum (peak of the electron beam).  For the different times $t=2.1, 3.2, 6.3$~s the positions are $r=0.25, 0.39, 0.75~R_\odot$, respectively.

% An initial increase in the maximum electron velocity that resonates with Langmuir waves has been shown previously in \citet{Kontar:2001ac} for a initial beam that has a Maxwellian distribution in velocity space.  Previous numerical studies \citep[e.g.][]{Reid:2013aa,Ratcliffe:2014aa,Li:2014aa,Khalilpour:2015aa} have shown that the velocity range of electrons which generate the bulk of the Langmuir waves decreases as a function of time.  The decrease in type III drift rate as a function of frequency \citep[e.g.][]{Fainberg:1972aa} has, in part, been attributed to the decrease in the resonant velocities as a function time.  

\begin{figure}\center
\includegraphics[width=\wfig,trim=40 0 20 18,clip]{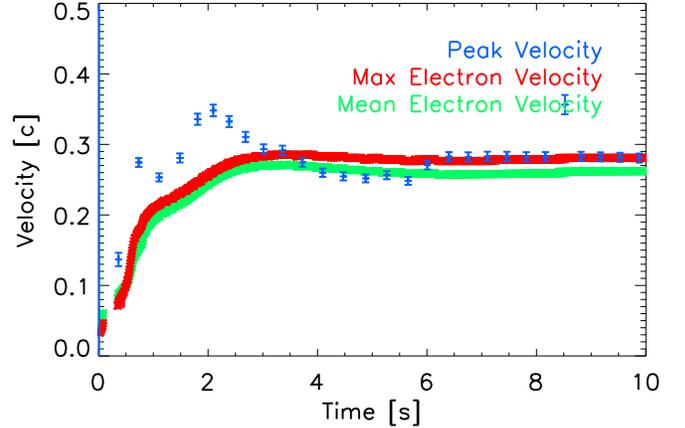}
\caption{Peak velocity of the electron beam from Figure \ref{figure:vel_time} over-plotted on the maximum and mean velocities in the electron distribution function as a function of time.  The peak velocity appears to track well the maximum electron velocity in the plateau at the same point in space.}
\label{figure:dist_time_f}
\end{figure}

The motion of electrons and Langmuir waves can be described as a beam-plasma structure using gas-dynamic theory \citep[e.g.][]{Ryutov:1970aa,Kontar:1998aa,MelNik:1999ab}.  Electrons are able to fully relax to a plateau in velocity space then the beam-plasma structure moves through space with the mean velocity of the electrons $v_{\rm bp}=(v_{\rm max}+v_{\rm min})/2$, where $v_{\rm max}$ and $v_{\rm min}$ are the minimum and maximum velocities within the plateau.  If the electrons have not fully relaxed then the mean electron velocity within the beam can be found using
\begin{equation}
\bar v(r,t) = \frac{\int_{v_{\rm min}}^{v_{\rm max}}{f(v,r,t)vdv}}{\int_{v_{\rm min}}^{v_{\rm max}}{f(v,r,t)dv}}.
\end{equation}
We show in Figure \ref{figure:dist_time_f} both the mean electron velocity and the velocity where the electron distribution $f(v)$ has a maximum, at the position corresponding to the peak brightness temperature, as a function of time.  We also show the velocity derived from the motion of the peak brightness temperature.

At the earliest times before, before three seconds, the peak velocity is significantly higher than either the maximum or mean velocity of the electrons which form the plateau.  As explained above, this is because the peak velocity is tracking the position of the peak brightness temperature.   Indicated in Figure \ref{figure:f_vel_time}, this position increases with time not just because of electron movement through space but also because the related electron beam is at higher (and then lower) velocities at different points in time.  The apparent motion is thus faster than the velocity of the electrons generating the Langmuir waves.

At the later times the peak velocity is similar to the velocity of the electrons resonating with the Langmuir waves.  There is a tendency to propagate at the same velocity as the maximum velocity of electrons within the plateau at the same spatial location.  This is different than what is expected from gas-dynamic theory, which predicts the mean velocity of electrons would track the motion.  The anomalous point around 8.5 seconds is just related to the discretisation of the grid in velocity space, and the sharp, localised increase in velocity should be spread over a longer time frame.

\section{Electron beam parameters} \label{sec:params}

The number density of electrons at specific velocities (the phase-space density) of the injected electron beam has a significant effect in dictating how fast the electron beam will travel through the coronal and interplanetary plasma.  Typically, the more high-energy electrons in the beam, the faster it will go.  Two key beam parameters that dictate this are the initial spectral index $\alpha$ and the initial beam density $n_b$.  

In this section we will explore how the velocity of the front, peak and back of the electron beam ($v_{\rm f}, v_{\rm p}, v_{\rm b}$, respectively) are affected by $\alpha$ and $n_b$.  To estimate how they affect beam propagation we varied the initial spectral index such that $6 \leq \alpha \leq 10$ and the initial beam density such that $10^{-1.5} \leq n_b/n_e \leq 10^{-3.5}$, where $n_e=3\times10^{9}~\rm{cm}^{-3}$ at the injection site.  The rest of the beam parameters are the same as Section \ref{sec:dynamics} and are given in Table \ref{tab:beam_sun}.

\begin{figure*}\center
\includegraphics[width=.49\textwidth,trim=30 0 20 38,clip]{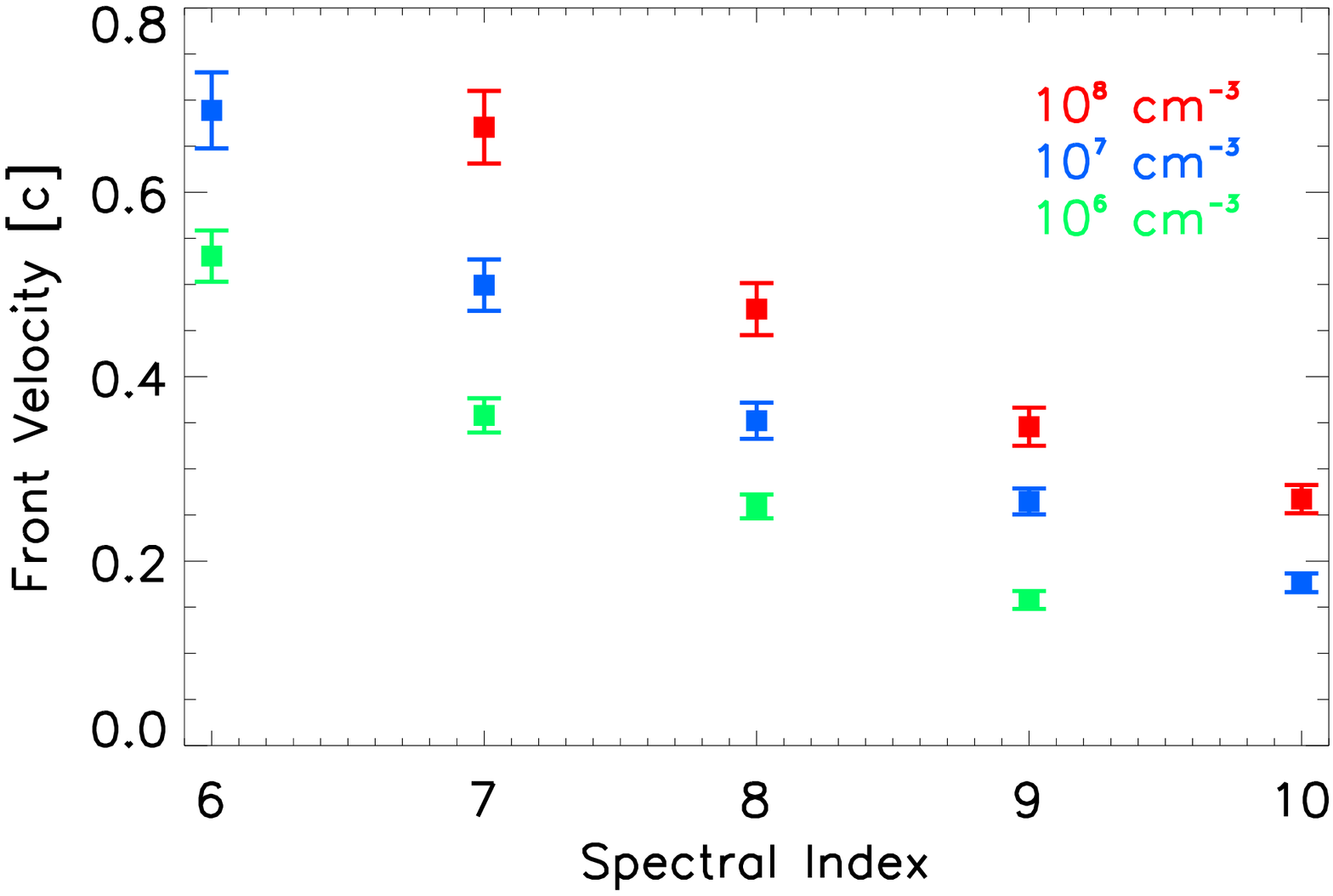}
\includegraphics[width=.49\textwidth,trim=30 0 20 38,clip]{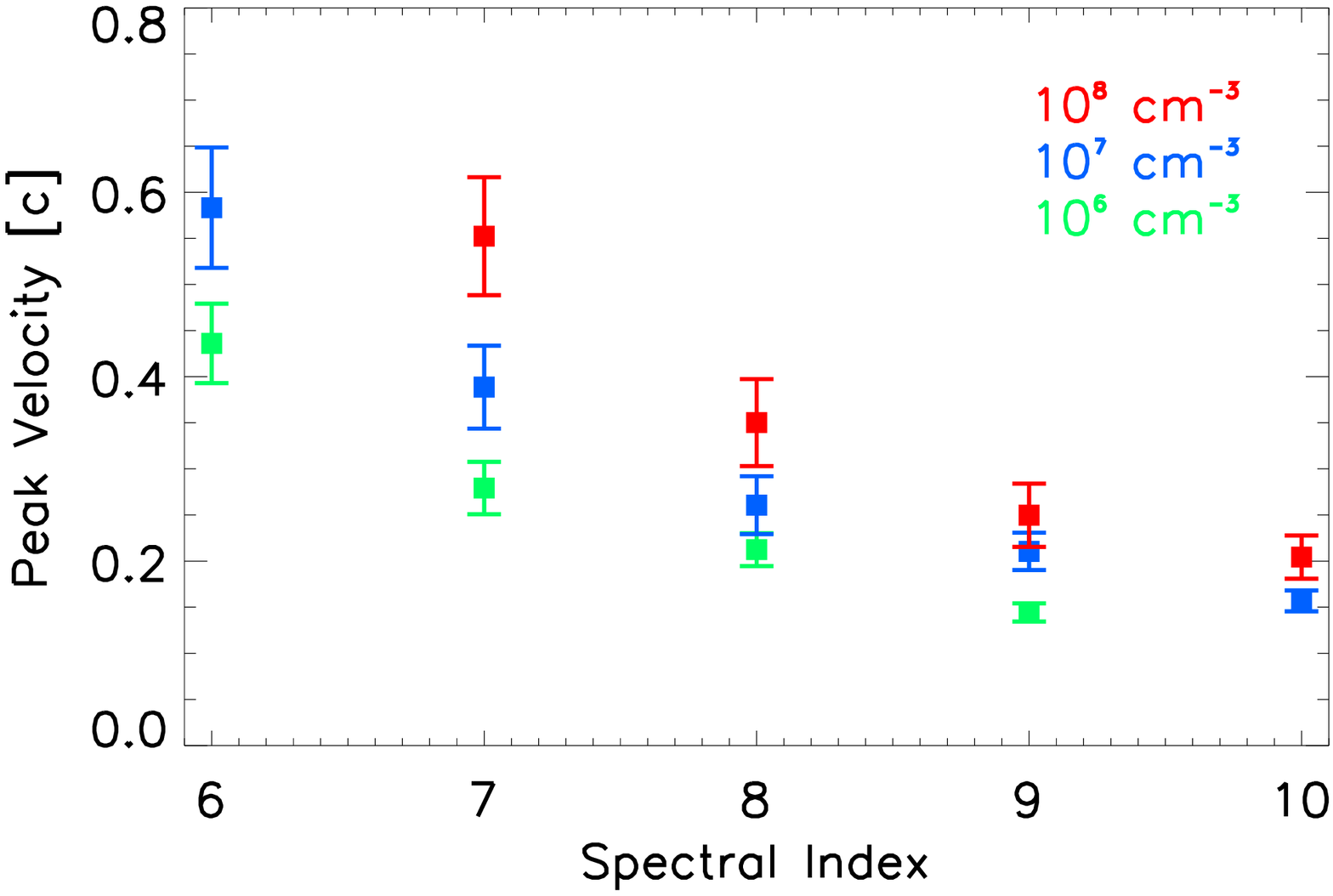}
\includegraphics[width=.49\textwidth,trim=30 0 20 38,clip]{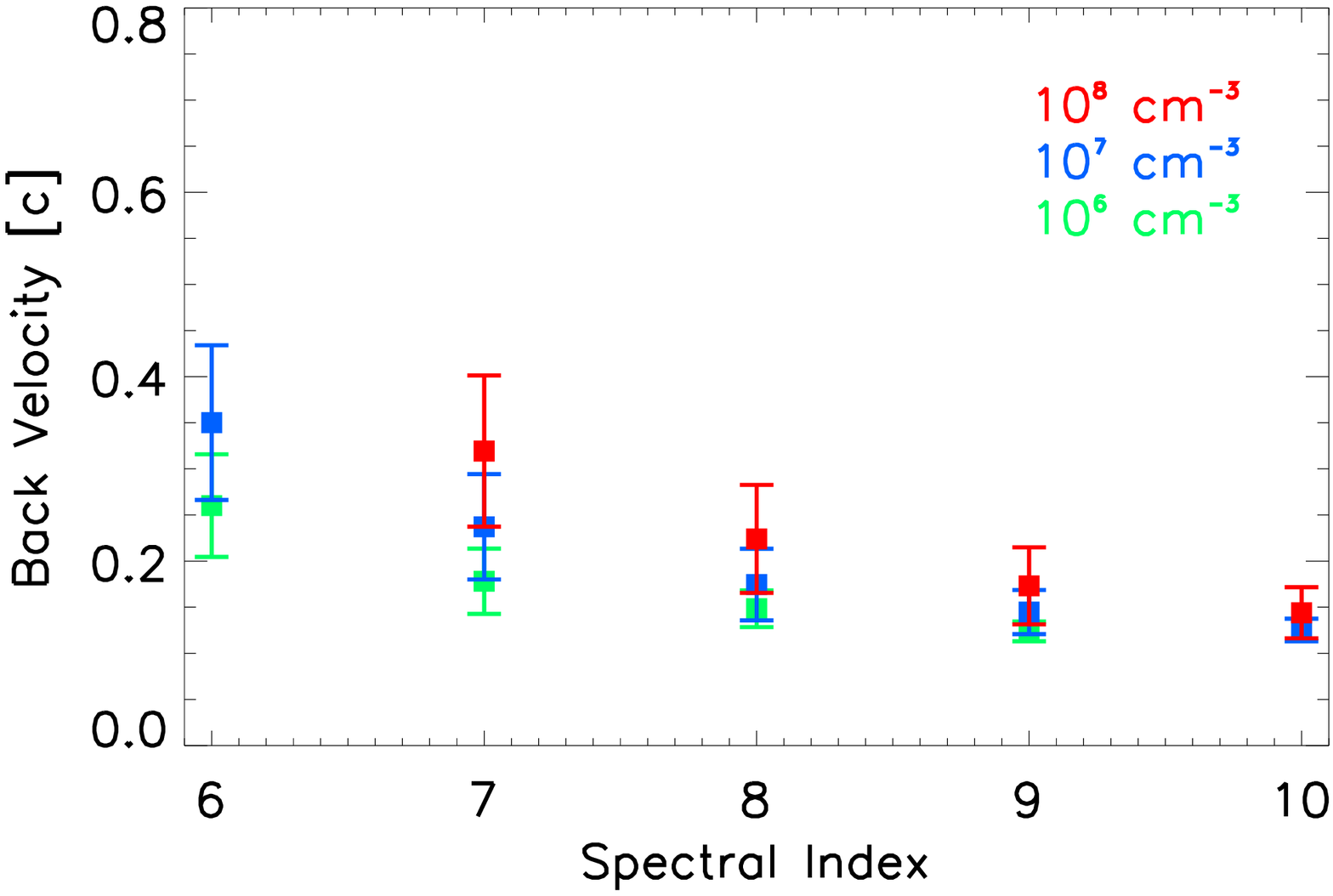}
\includegraphics[width=.49\textwidth,trim=30 0 20 38,clip]{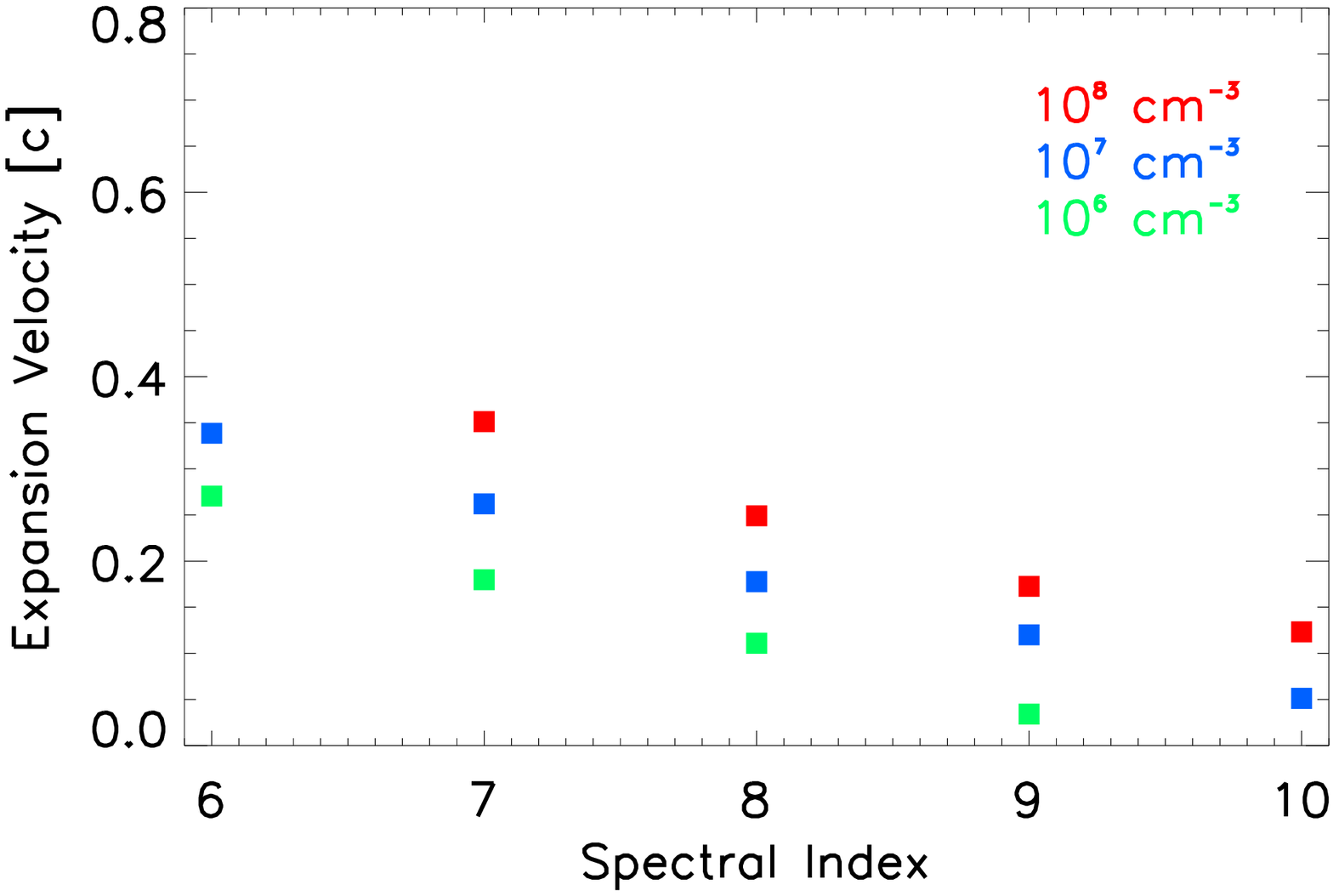}
\caption{Apparent velocity of the front, peak, back and expansion of the beam as a function of spectral index for beam densities $10^6, 10^7, 10^8~\rm{cm}^{-3}$.  The velocities are found using a linear fit of distance and time between 30--70~MHz.  The width of the error bars indicate the mean width of the electron distribution function when the front, peak and back velocities are calculated.}
\label{figure:si_vel}
\end{figure*}

As shown in Figure \ref{figure:vel_time} the velocity of the electron beam varies as a function of distance.  To compare simulations with a single velocity, we find the average velocity over the distance range that corresponds to plasma frequencies 30--70 MHz.  These frequencies relate to distances $3.0\times10^{10}-5.3\times10^{10}~\rm{cm}$ or $0.44-0.76~R_\odot$.  We find the average velocity from the gradient of a linear fit to distance as a function of time.

\subsection{Electron beam velocities}

The velocity at the front of the electron beam is shown in Figure \ref{figure:si_vel} for different values of $\alpha$ and $n_b$.  The velocity increases as the spectral index decreases and the initial beam density increases.  Both parameters show that increasing the number of electrons at  velocities higher than the thermal speed controls how fast the front of the electron beam moves.  The front velocity, $v_{\rm f}$, when $n_b=10^7~\rm{cm}^{-3}$ and $\alpha=6$ is slightly lower than we might expect because substantial Langmuir waves are generated by electrons with velocities $v_{\rm max}=0.7c$, and would likely be generated by higher velocity electrons, if included in the simulation.  

The peak velocity, $v_{\rm p}$ of the electron beam changes as the initial density and spectral index of the electron beam is altered, shown in Figure \ref{figure:si_vel}.  Similar to $v_{\rm f}$, $v_{\rm p}$ increases for smaller initial spectral indices and larger initial beam densities.  The size of the peak velocities are smaller than the front velocities, such that the mean $\pm$ the standard deviation of $v_{\rm p}/v_{\rm f}$ was $0.80\pm0.06$.

The velocity of the back of the beam, $v_{\rm b}$,  varies in a similar manner, increasing with decreasing initial spectral index and increasing initial beam density, shown in Figure \ref{figure:si_vel}.  The back velocity is smaller such that $v_{\rm b} < v_{\rm p} < v_{\rm f}$.  The mean ratio of $v_{\rm b}/v_{\rm p}$ was $0.67\pm0.08$ whilst the mean ratio of $v_{\rm p}/v_{\rm f}$ was $0.54\pm0.1$.  The variation in the back velocity as the initial beam parameters are changed is much less than the front and peak velocities.  The back velocity is heavily dependent on the thermal velocity of the background Maxwellian plasma; we investigate this dependence in the next section.

The mean width of the electron distribution function when the front, peak and back velocities are calculated are displayed using error bars in Figure \ref{figure:si_vel}.  The width, $dv$, is defined using the full width, $10\%$ maximum on account of the sharp decrease in the distribution function afterwards (see Figure \ref{figure:f_vel_time}).  Defining $dv$ using the FWHM reduces the width by 2, or 1.5 for the back velocities.  The ratio $dv/v$ gives an indication of how much quasilinear diffusion the beam has undertaken.  Mean ratios for the front, peak and back of the beam are $10\%$, $20\%$, $40\%$, respectively.  The low value for $dv/v$ indicates that the beams are only weakly relaxing, in comparison to the full plateau that is assumed using gas-dynamic theory.  The mean ratio increases towards the back of the beam on account of the higher beam densities at lower velocities making the quasilinear diffusion quicker, and hence more pronounced.

The change in beam length as a function of time, the expansion velocity $\Delta v \approx v_{\rm f}-v_{\rm b}$, is shown in Figure \ref{figure:si_vel}.  It is not surprising that we find $\Delta v$ having similar behaviour to $v_{\rm f}$ and $v_{\rm b}$, increasing with higher initial beam densities and lower initial spectral indices.

\begin{figure}\center
\includegraphics[width=\wfig,trim=30 0 20 38,clip]{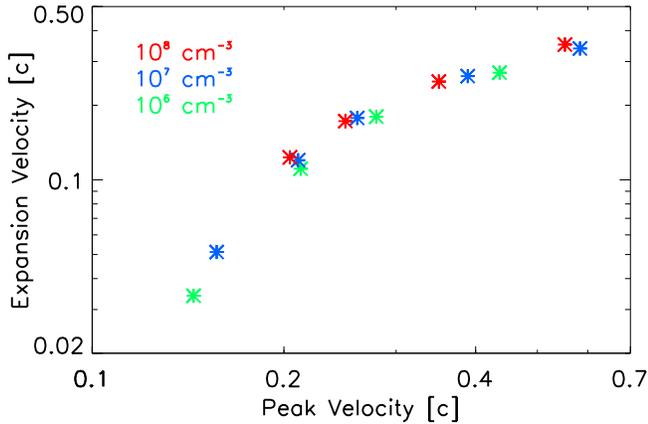}
\caption{The expansion velocity as a function of peak velocity, using results shown in Figures \ref{figure:si_vel}.}
\label{figure:vel_peak_length}
\end{figure}

The expansion velocity has a linear correlation with the peak velocity, shown in Figure \ref{figure:vel_peak_length}, with a Pearson correlation coefficient of 0.98.  As the peak velocity increases, the expansion velocity increases.  The expansion velocity tends to zero around $v_{\rm peak}=0.1~c$.  Around this velocity, Landau damping from the 1~MK background plasma suppresses any wave generation by the electron beam; the reason why $v_{\rm peak}$ does not go any lower.

\subsection{Beam velocity evolution}

In the previous section we investigated beam velocities by averaging averaged between 70--30~MHz to obtain a single velocity, for comparison purposes.  However, the beam velocity is not constant with either time or space, as shown in Figure \ref{figure:vel_time}.  It varies, depending on the minimum and maximum electron velocities that are important for the beam-plasma interaction with Langmuir waves.

\begin{figure}\center
\includegraphics[width=\wfig,trim=30 0 20 38,clip]{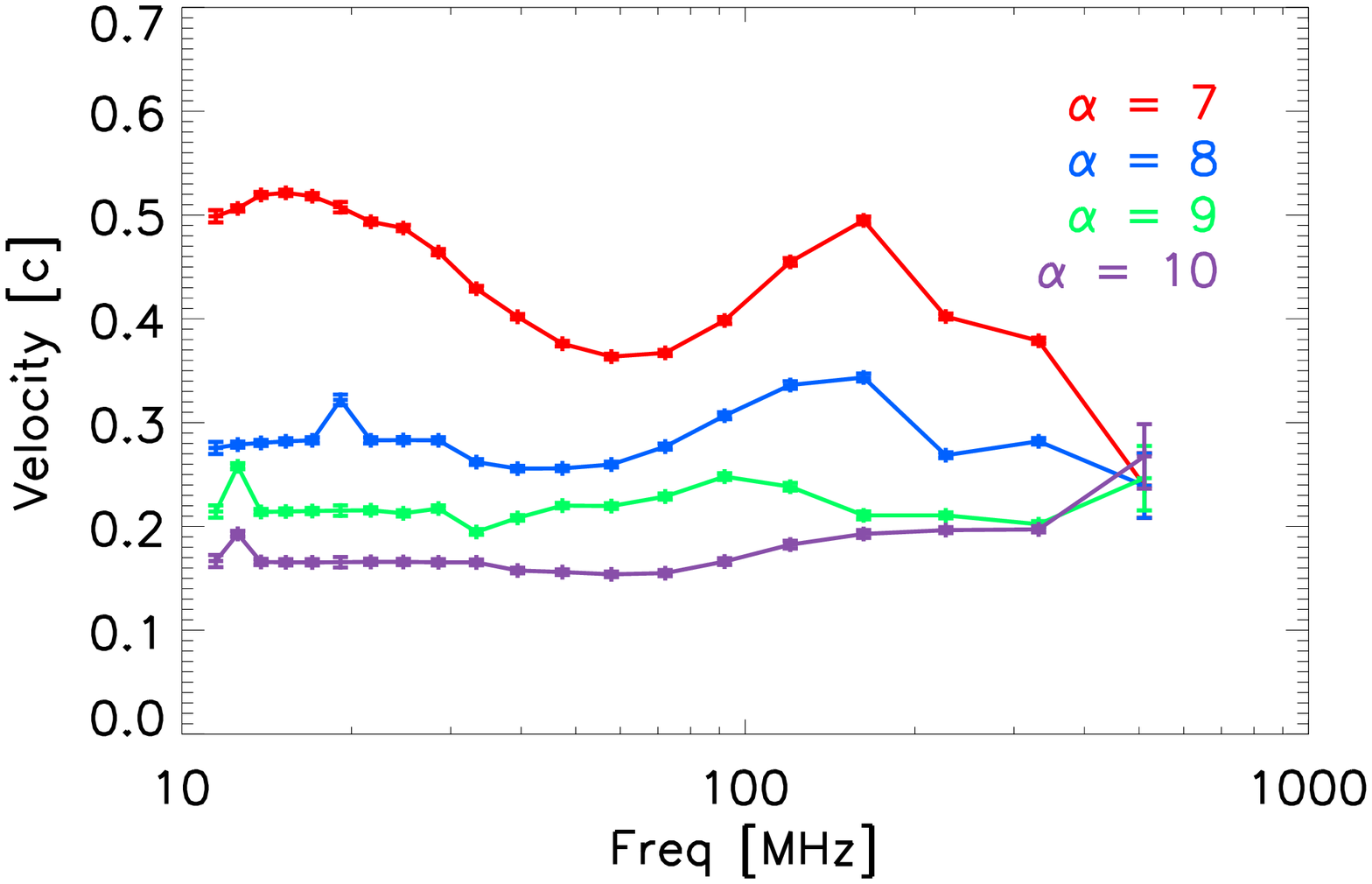}
\includegraphics[width=\wfig,trim=30 0 20 38,clip]{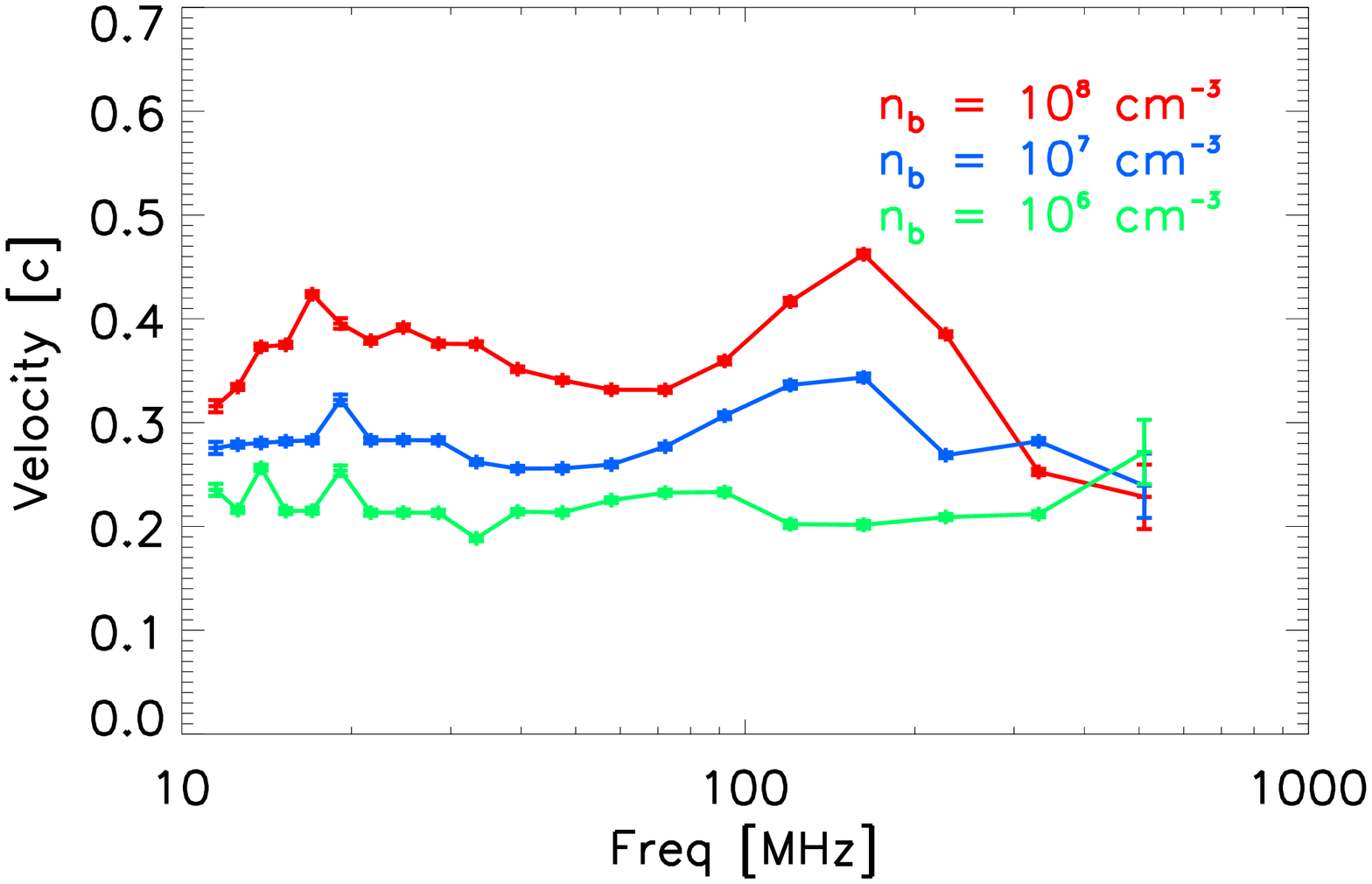}
\caption{Beam velocity as a function of plasma frequency.  Top: the initial spectral index $\alpha$ varies from 7 to 10 with an initial beam density of  $n_b=10^{7}~\rm{cm}^{-3}$.  Bottom: the initial beam density varies from $n_b=10^{6}~\rm{cm}^{-3}$ to $n_b=10^{8}~\rm{cm}^{-3}$ with an initial spectral index of $\alpha =8$.}
\label{figure:vel_freq}
\end{figure}

We show in Figure \ref{figure:vel_freq} the velocity evolution of the peak brightness temperature as a function of frequency for  simulations with different initial spectral indices and beam densities.  The increase and subsequent decrease in beam velocity at frequencies above 100~MHz is consistent with all simulations, with the explanation given in Section \ref{sec:vel_evolve}.  

At frequencies below 100~MHz, we observe an increase in the velocity of the peak brightness temperature for some simulations but not others.  For the simulations where the number of high-energy electrons (at deca-keV energies) is higher, due to a higher beam density or a lower spectral index, the velocity tends to increase and then decrease.  For example, when $\alpha=7$ and $n_b=10^{7}~\rm{cm}^{-3}$, there are enough deca-keV electrons that higher and higher electron velocities become relevant to the beam-plasma structure with distance.  This occurs until such point where the decrease in beam density from the expanding flux tube starts making wave generation more difficult.  Lower and lower electron velocities then become more relevant and the velocity of the peak brightness temperature decreases.   For the simulations where, e.g, $\alpha=10$ and $n_b=10^{7}~\rm{cm}^{-3}$, this point occurs before 100 MHz and we do not see a significant change in the velocity.  There are some anomalous points in Figure \ref{figure:vel_freq} at frequencies below 20~MHz, again related to the discretisation of the grid in velocity space, and the sharp, localised increase in velocity should be spread over a wider frequency band.

An increase in the inferred beam velocity from type III bursts has recently been reported by \citet{Mann:2018aa} using LOFAR imaging observations below 100 MHz.  \citet{Mann:2018aa} derive velocities from the onset time of the type III burst at different frequencies and found huge increases in derived beam velocities around 30~MHz, with some bursts reported as superluminal.  Whilst the effect we describe above can increase the derived beam velocity from radio bursts, and could play a role in the findings of \citet{Mann:2018aa}, it is not likely to explain three-fold increases in velocities, nor provide superluminal velocities.  Whilst refraction effects are considered, radio wave scattering at LOFAR frequencies has a more dominant effect \citep[see][]{Kontar:2017ab}, is likely to increase the derived beam speeds and could explain the apparent beam acceleration.

\begin{figure}\center
\includegraphics[width=\wfig,trim=30 0 10 38,clip]{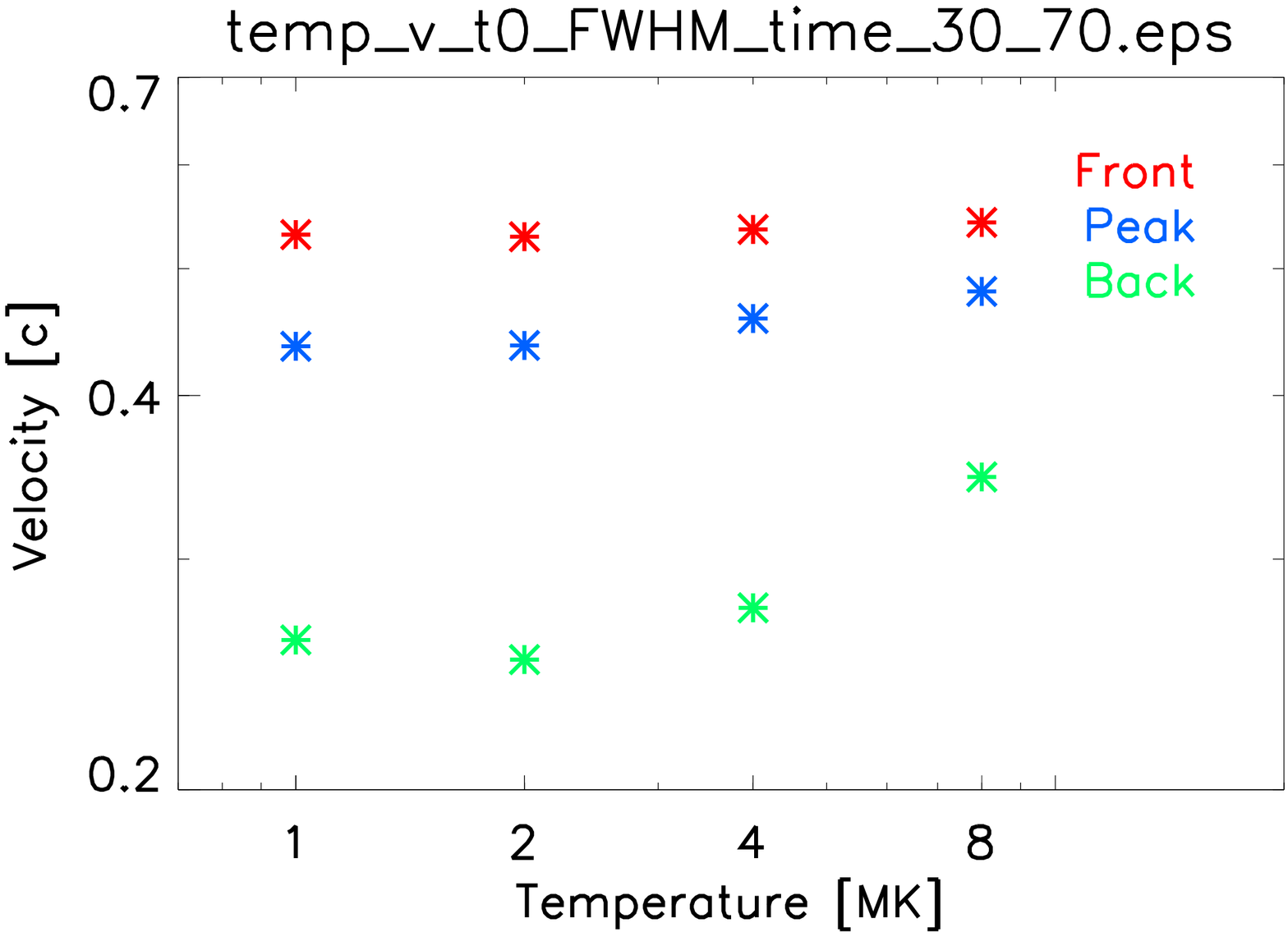}
\caption{Front, peak and back velocities for an electron beam travelling through background plasma with a temperature of 1, 2, 4, 8 MK.  Velocities are found using a linear fit of distance and time between 20--35 MHz, corresponding to 40--70~MHz assuming harmonic emission.}
\label{figure:temp_v}
\end{figure}

\begin{figure*}\center
\includegraphics[width=.49\textwidth,trim=550 25 0 0,clip]{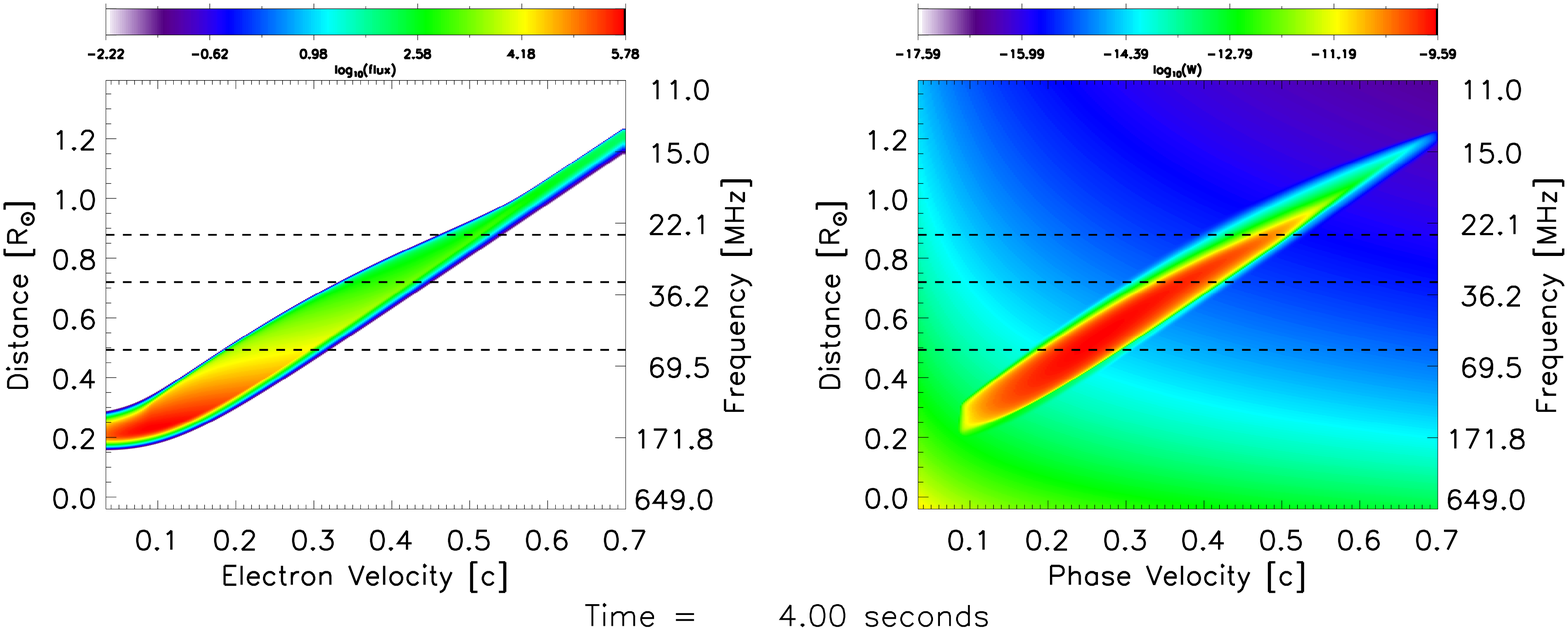}
\includegraphics[width=.49\textwidth,trim=550 25 0 0,clip]{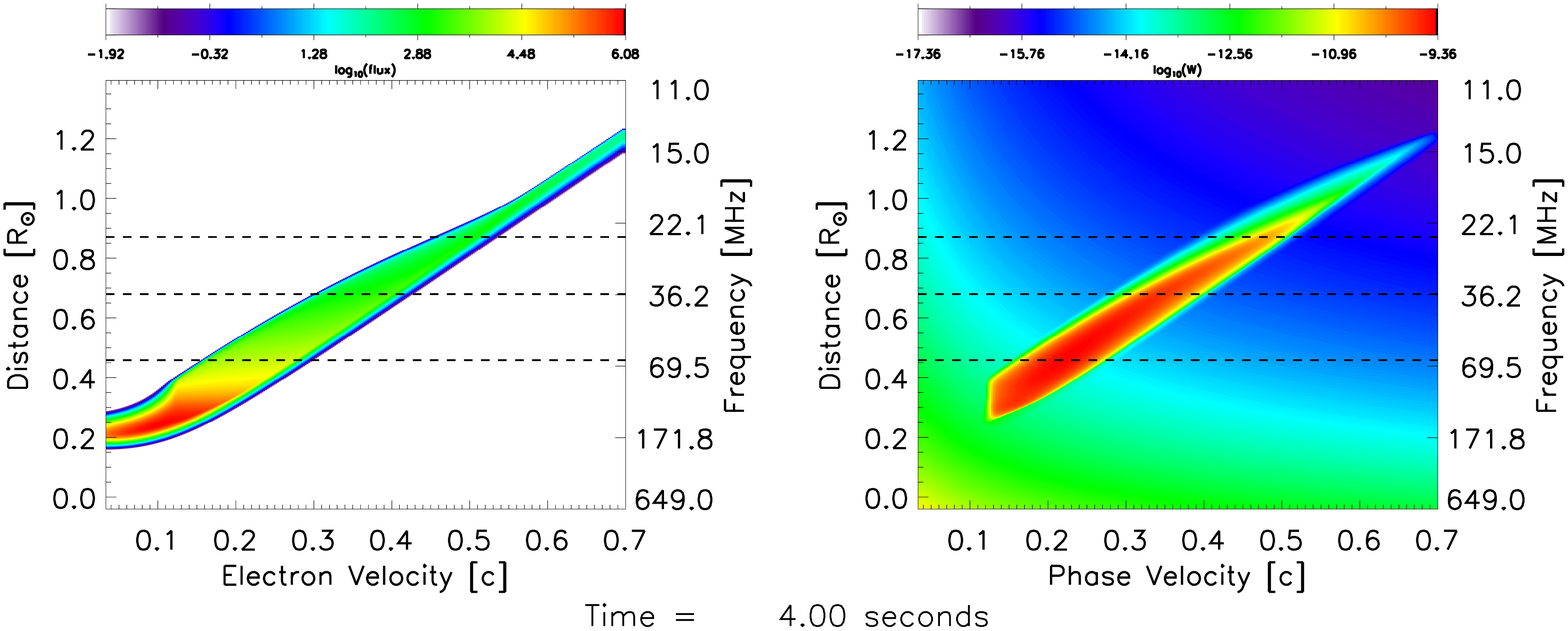}
\includegraphics[width=.49\textwidth,trim=550 25 0 0,clip]{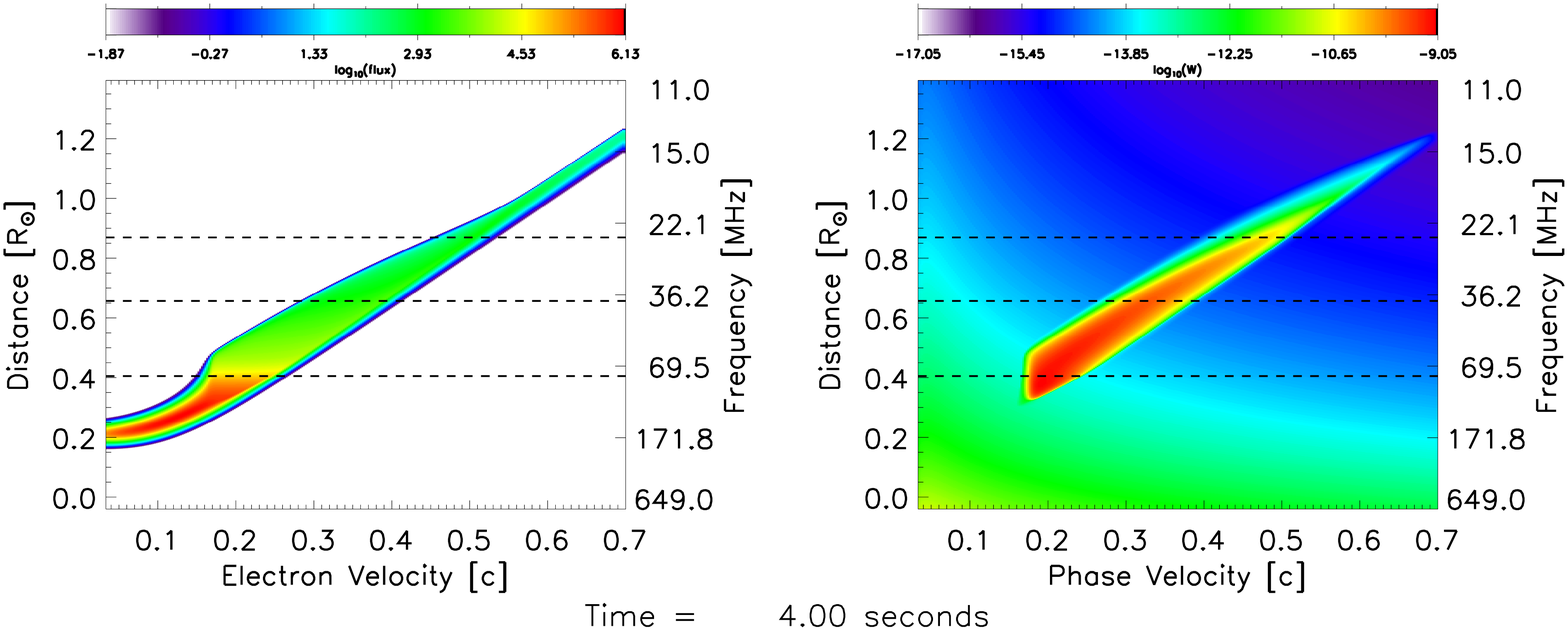}
\includegraphics[width=.49\textwidth,trim=550 25 0 0,clip]{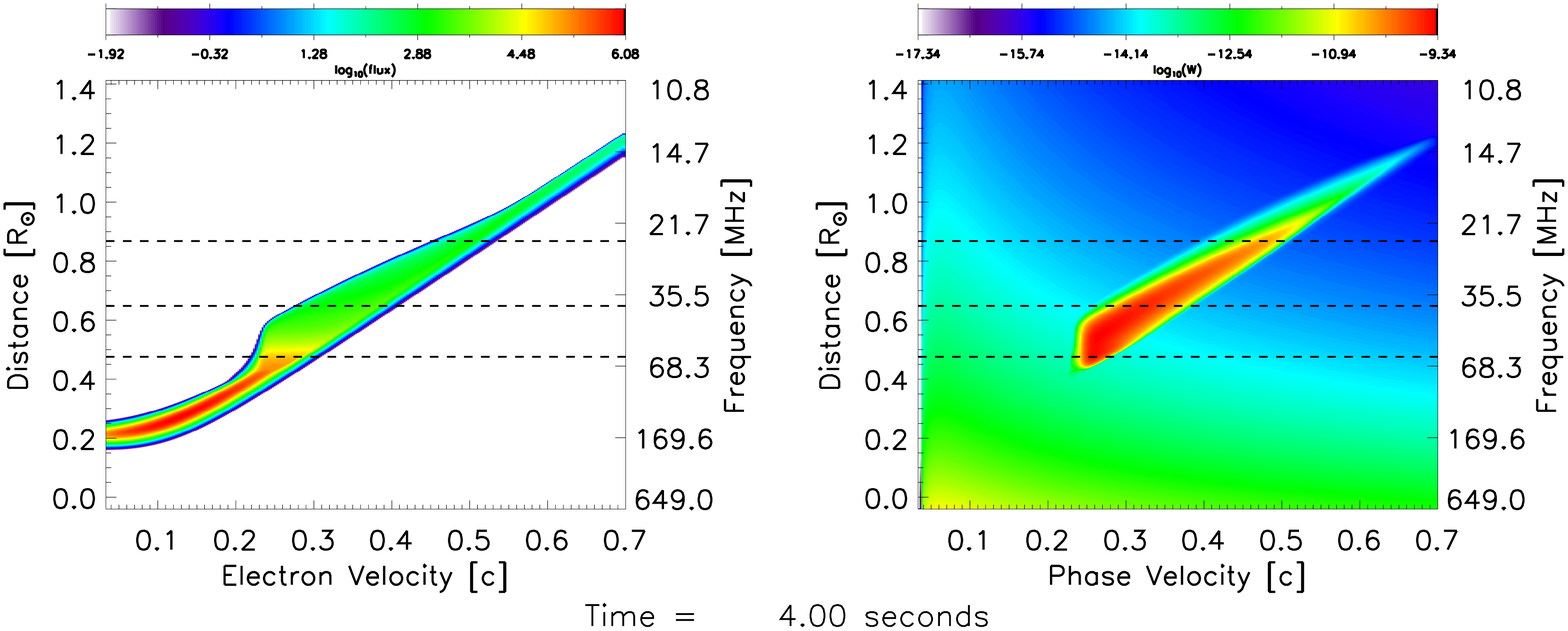}
\caption{Langmuir wave spectral energy density after 4 seconds of beam propagation.  The background temperature varies as 1, 2, 4, 8 MK from top left to bottom right.  The front, peak and back of the electron beam, found from the resulting radio brightness temperature, are shown by the horizontal dashed lines.  Note the decrease in Langmuir waves at lower phase velocities for higher background temperatures.}

\label{figure:temp}
\end{figure*}

\subsection{Thermal Velocity}

The background plasma temperature plays a significant role in governing the largest velocity at which Langmuir waves are damped by the background plasma.  Langmuir waves are heavily Landau damped close to the thermal velocity.  Our previous simulations used a background temperature of 1~MK, giving a thermal velocity of $v_{te}=3.9\times10^8~\rm{cm~s}^{-1}$ or 0.013~c.  To show the effect of the background temperature, Figure \ref{figure:temp} shows a snapshot of the Langmuir wave spectral energy density for four different simulations using the same initial electron beam parameters ($\alpha=6, n_b=10^6~\rm{cm}^{-3}$) but with different background temperatures of $1, 2, 4, 8$~MK.  This corresponds to $v_{te}=0.013, 0.018, 0.026, 0.037$c, respectively.  It is clear from Figure \ref{figure:temp} that a higher background temperature increases the minimum velocity where a substantial level of Langmuir waves are generated.

Higher background temperatures increase the velocities of the back of the electron beam, shown in Figure \ref{figure:temp_v}.  This is because slower electrons are no longer able to generate substantial levels of Langmuir waves in the presence of high Landau damping.  For temperatures of 1~MK, the back of the beam relates to Langmuir waves with phase velocities around 0.25c, well above the lowest phase velocities of the Langmuir waves.  At temperatures of 8~MK, the back of the beam relates to Langmuir waves with the lowest phase velocities.  This behaviour is captured by the ratio of $v_{\rm b}/v_{te}$ that is 20 for $T_e=1$~MK and 9.4 for $T_e=8$~MK.  Conversely, the front of the beam is not influenced very much by the background temperature as the corresponding Langmuir waves at high phase velocities are not damped by the background electron distribution.

\section{Type III bandwidth and drift rate} \label{sec:band_drift}

\subsection{Type III bandwidth}

The instantaneous bandwidth of a type III burst is the width in frequency space at any one point in time.  In the context of the electron beam, this relates to the range of plasma frequencies within the beam length; the difference in plasma frequency between the position of the front and the back of the electron beam.

\begin{figure}\center
\includegraphics[width=\wfig,trim=30 0 10 38,clip]{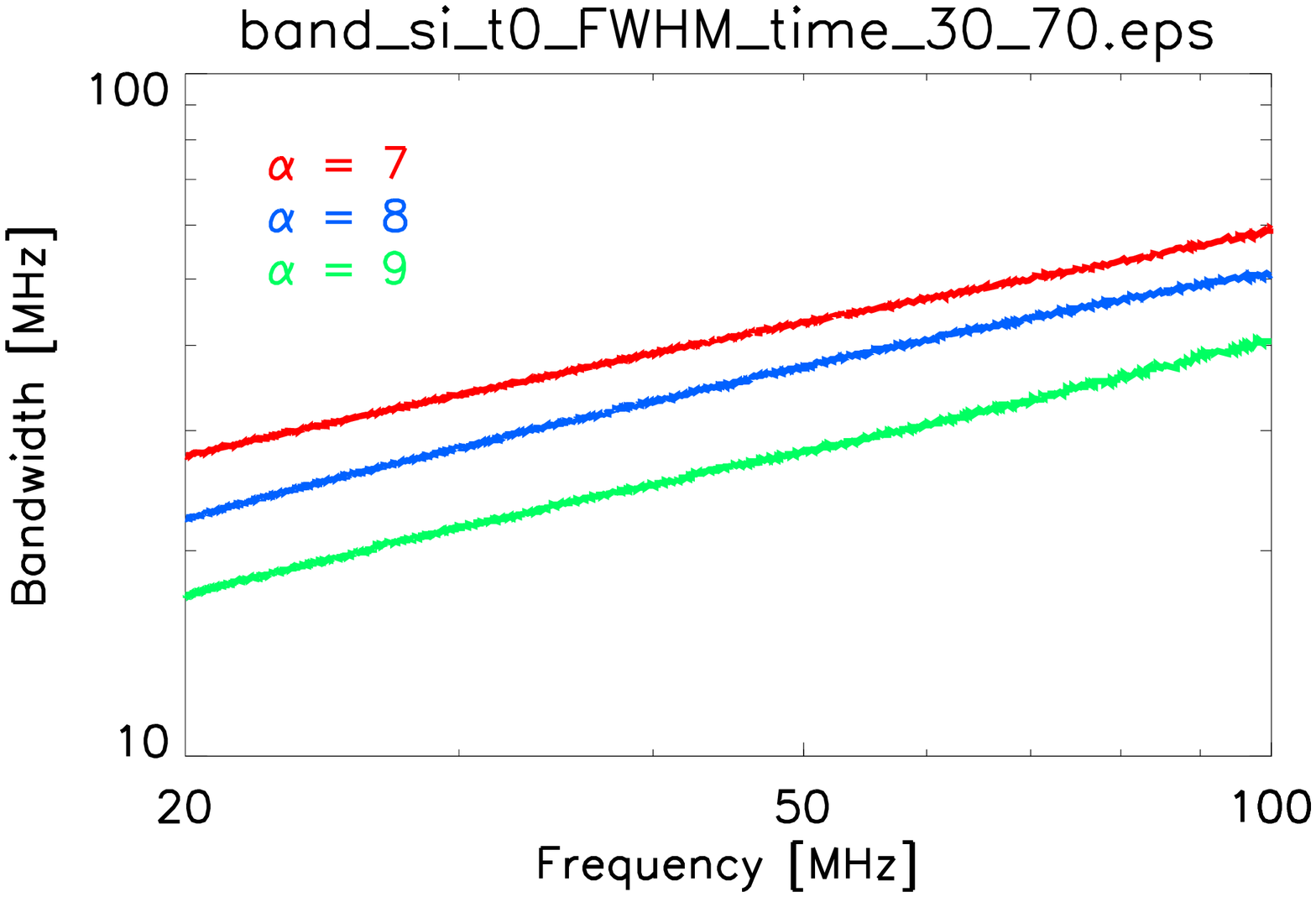}
\includegraphics[width=\wfig,trim=30 0 10 38,clip]{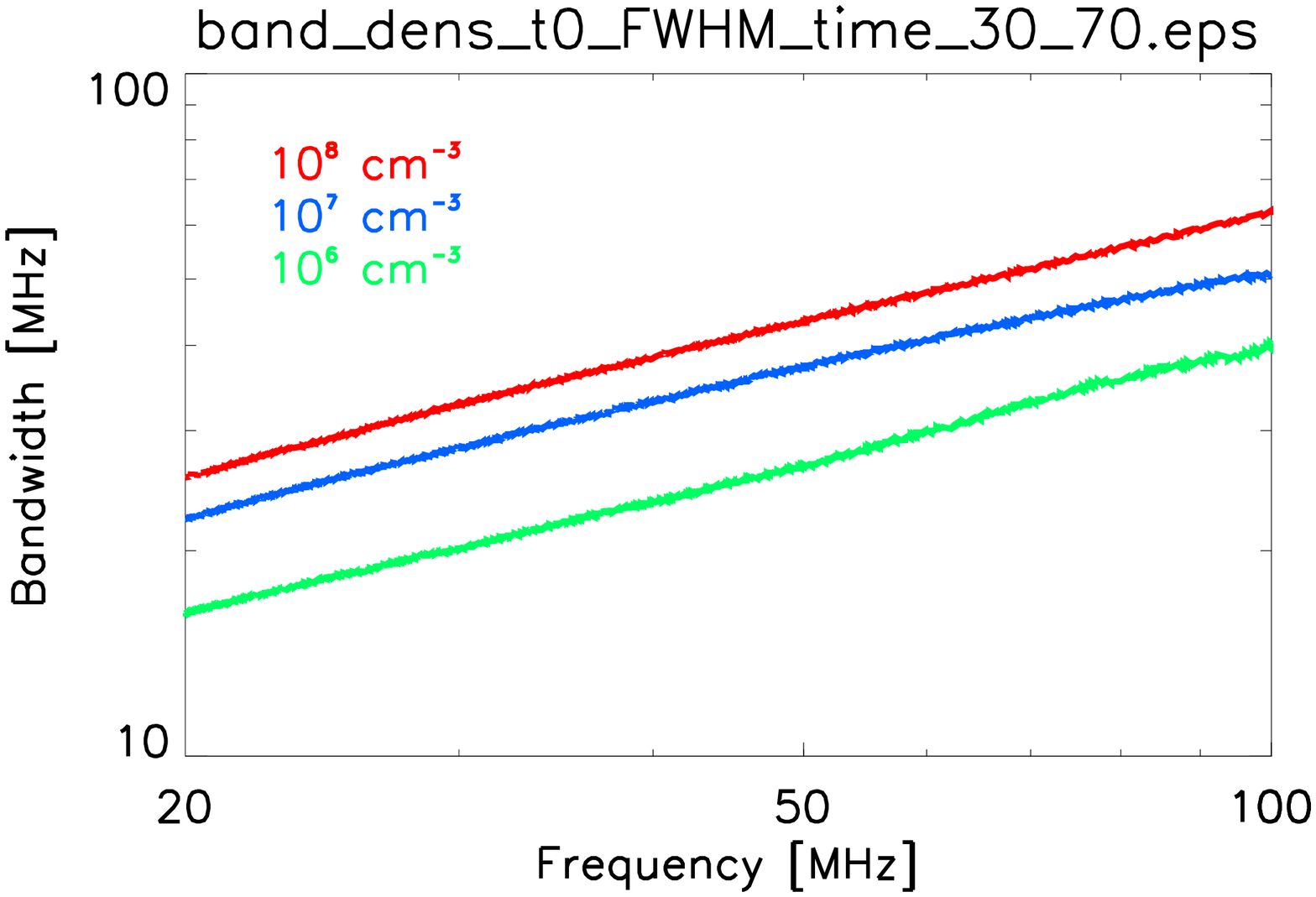}
\caption{Instantaneous bandwidth of the electron beam as a function of the plasma frequency.  Top: the initial spectral index $\alpha$ varies from 7 to 9 with an initial beam density of  $n_b=10^{7}~\rm{cm}^{-3}$.  Bottom: the initial beam density varies from $n_b=10^{6}~\rm{cm}^{-3}$ to $n_b=10^{8}~\rm{cm}^{-3}$ with an initial spectral index of $\alpha =8$.}
\label{figure:sim_bandwidth}
\end{figure}

Figure \ref{figure:sim_bandwidth} shows how the bandwidth is different for different initial spectral indices.  The frequency associated with each bandwidth is the background plasma frequency at the location of the peak of the electron beam.  For all frequencies, the bandwidth is higher for lower initial spectral indices.  This is expected from Figure \ref{figure:si_vel} as a lower initial spectral index creates a beam with a longer length.  There is also a systematic decrease in the bandwidth as the frequency decreases.  This is related to the bulk decrease in magnitude of the background electron density gradient.  Over nearly an order of magnitude in frequency the change in bandwidth from the spectral index is reasonably significant, with the bandwidth at 100~MHz at $\alpha=9$ being the same as the bandwidth at 40~MHz at $\alpha=7$.

The bandwidth is affected by the initial beam density, shown in Figure \ref{figure:sim_bandwidth}, with a higher initial beam density increasing the bandwidth.  Again the change in bandwidth from the initial beam density is significant in comparison to the change in bandwidth from the decrease in magnitude of the background electron density gradient at lower frequencies.

\subsection{Type III drift rate}

\begin{figure}\center
\includegraphics[width=\wfig,trim=30 0 10 38,clip]{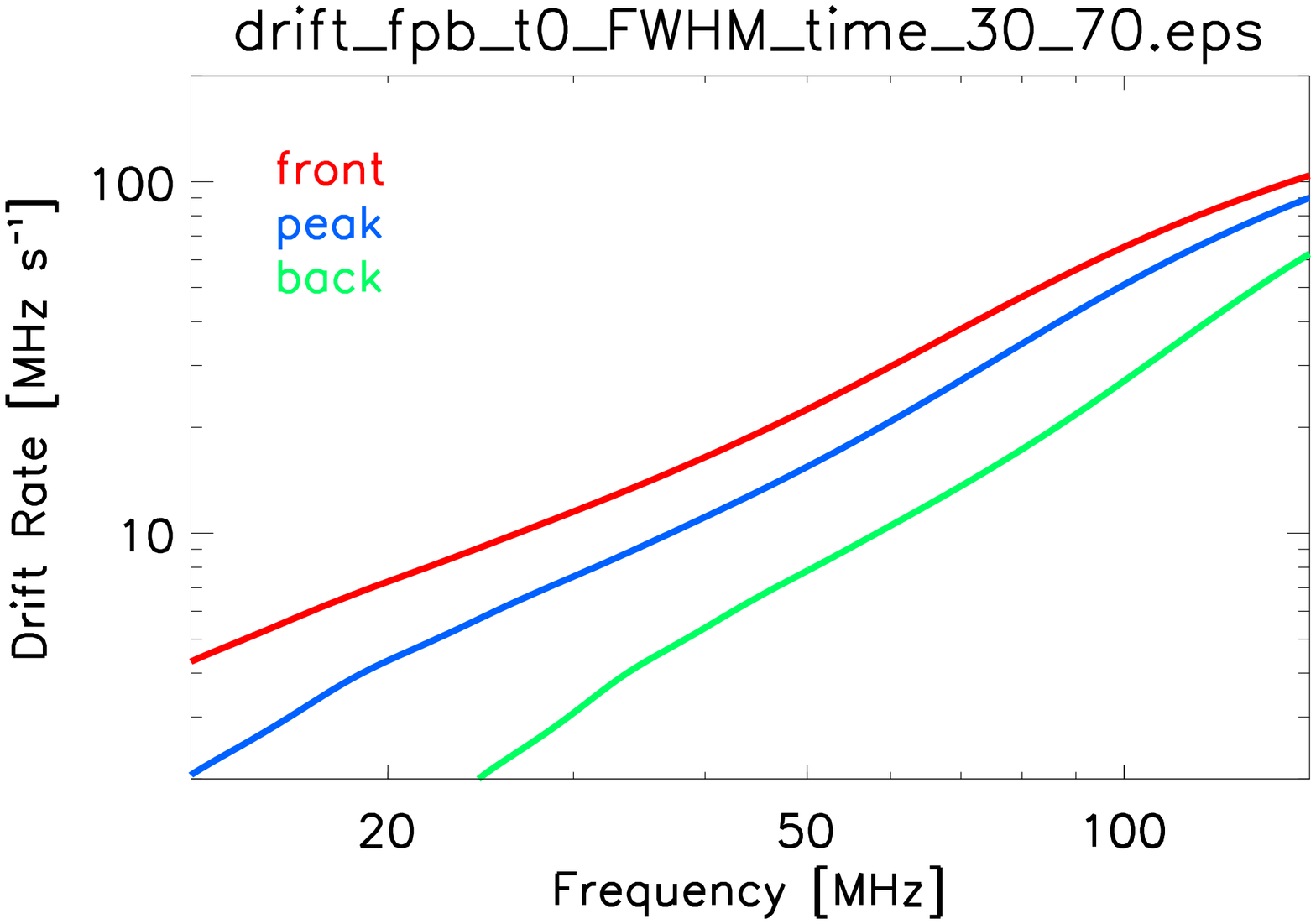}
\caption{Frequency drift rate magnitude of the rise, peak and decay of the type III burst as a function of frequency.  This related to the change in position of the front, peak and back of the beam as a function of time.  The initial beam density was $n_b=10^{7}~\rm{cm}^{-3}$ and the initial spectral index was $\alpha=8$.}
\label{figure:dfdt_fpb}
\end{figure}

\begin{figure}\center
\includegraphics[width=\wfig,trim=30 0 10 34,clip]{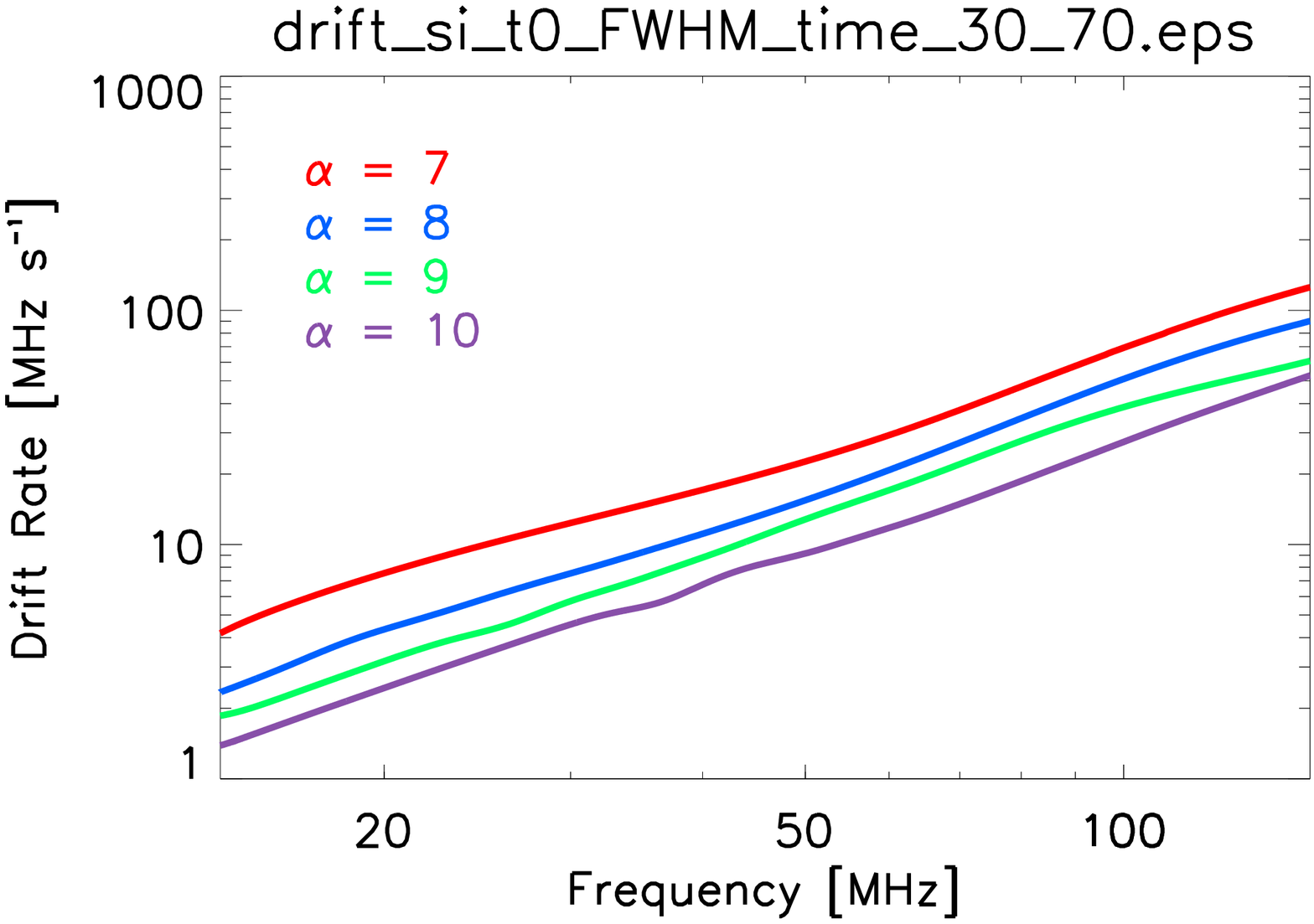}
\includegraphics[width=\wfig,trim=30 0 10 38,clip]{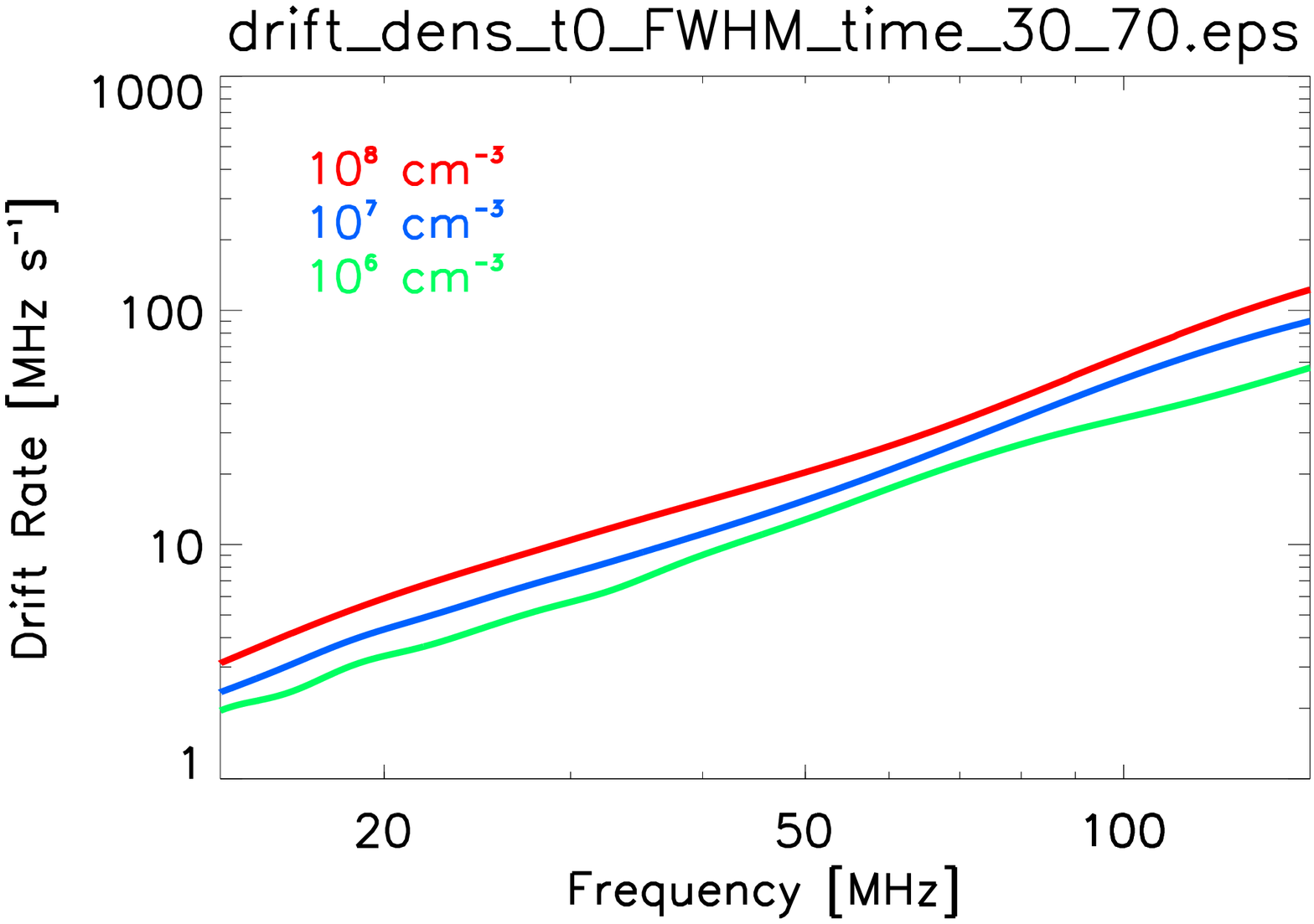}
\caption{Frequency drift rate magnitude of the electron beam as a function of plasma frequency.  Top: the initial spectral index $\alpha$ varies from 7 to 10 with an initial beam density of  $n_b=10^{7}~\rm{cm}^{-3}$.  Bottom: the initial beam density varies from $n_b=10^{6}~\rm{cm}^{-3}$ to $n_b=10^{8}~\rm{cm}^{-3}$ with an initial spectral index of $\alpha =8$.}
\label{figure:dfdt}
\end{figure}

The drift rate of type III bursts can be found using the variation of frequency with time at a given intensity point (e.g. time of peak intensity in each frequency channel), found using the rise, peak or decay times in the type III dynamic spectra;  relating to the motion of the front, peak or back of the electron beam as a function of time.  Figure \ref{figure:dfdt_fpb} shows the drift rate inferred from the rise, peak and decay times of the fundamental emission calculated from the Langmuir wave spectral energy density.  The large difference in magnitude between the front and back drift rates is evident, with the front clearly drifting faster than the back, as expected.

The initial parameters also govern the magnitude of the drift rate as a function of frequency.  Figure \ref{figure:dfdt} shows how the inferred drift rate changes when the initial spectral index is varied.  Similar to the bandwidth, the magnitude of the drift rate decreases as the initial spectral index is increased.  Compared to bandwidth, the initial spectral index is less significant for the drift rate compared to the decrease in drift rate from the reduced background electron density gradient at lower frequencies.  For example the drift rate at 100~MHz when $\alpha=9$ is the same as the drift rate at 70~MHz when $\alpha=7$. 

The initial beam density also affects the drift rate, with a smaller initial beam density leading to a lower drift rate magnitude, shown in Figure \ref{figure:dfdt}.  Again, the change in drift rate from the initial beam density is less than the change in drift rate from the bulk decrease in background density density gradient at distances farther from the Sun.  For example the drift rate at 100~MHz when $n_b=10^6$ is the same as the drift rate at 70~MHz when $n_b=10^8$. 

The drift rates around 100 MHz are very similar to the drift rates derived from type III radio observations at similar frequencies by \citet{Alvarez:1973ab}.  However, a power-law fit to the peak curve in Figure \ref{figure:dfdt_fpb} gives a spectral index of 1.52, lower than \ the 1.84 found in \citet{Alvarez:1973ab}.  The drift rates are slightly higher than those given by \citet[][]{Achong:1975aa} between 26--36~MHz and by \citet{Reid:2018aa} between 30--70~MHz.  The spread in drift rates from the different electron beam parameters is similar to the spread in values presented in both studies.  Moreover, the spectral index of 1.52 is closer to those presented in the latter two studies, with a larger spectral index found from the back of the beam matching the results in \citet{Reid:2018aa}.

Using numerical simulations, a decrease in the magnitude of the drift rate has previously been shown by \citet{Li:2013aa} at 120~MHz and 80~MHz for an increasing initial power-law spectral index.  Similarly \citet{Kontar:2001ab} showed that decreasing the beam density or the characteristic velocity of a Maxwellian electron beam decreased the magnitude of the drift rate.  We highlight that the change in drift rate from the initial beam parameters does not significantly alter any type III drift rate as a function of frequency, compared to the effect of the background density model.  This is despite a doubling in the inferred peak velocity between spectral indices $\alpha=6$ and $\alpha=7$.

\section{Discussion} \label{sec:discussion}

We have shown that the initial electron beam parameters play a significant role on the dynamics of the electron beam as it travels through the heliosphere.  We initially assumed that the electron beam was a power-law in velocity space with a spectral index $\alpha$.  This lead to the spectral index and the beam density playing the dominant role for the phase space density of the electron beam.

\subsection{Initial Broken Power-law}

We can instead assume that the electron beam is injected as a broken power-law in velocity space with some break velocity $v_{\rm min} \leq v_0 \leq v_{\rm max}$.  The source function changes from $S(v,r,t) \propto v^{-\alpha}$ in Equation \ref{eqn:source} to 
\begin{equation}
S(v,r,t) \propto A_{v2}
	\begin{cases}
		1,& \text{if } v \leq v_0 \\
    	(v_0/v)^\alpha,& \text{if } v > v_0.
	\end{cases}
\end{equation}
where
\begin{equation}
A_{v2} = n_b \left[ v_0-v_{\rm min}+\frac{v_{\rm max}(v_{\rm max}/v_0)^{-\alpha} - v_0}{1-\alpha}\right]^{-1}.
\end{equation}
As the distribution below $v_0$ is flat, the number of electrons at velocities higher than a few $v_{\rm min}$ are much greater than Equation \ref{eqn:source} for a given beam density $n_b$.  The broken power-law is necessary to have very high densities contained within electrons above certain energies (e.g. 20~keV).  Very high densities above these energies are required to explain the hard X-ray observations observed in the low solar atmosphere \citep{Holman:2011aa}.  

\begin{figure}\center
\includegraphics[width=\wfig,trim=30 0 0 0,clip]{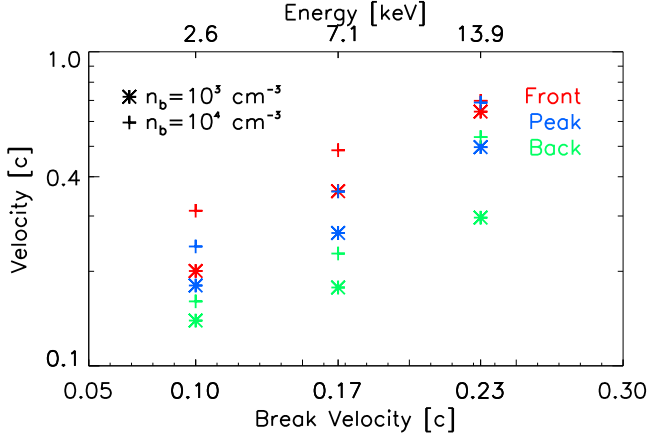}
\caption{Front, peak and back velocities for an electron beam with an initial broken power-law velocity distribution using different break velocities.  Different beam densities of $10^3, 10^4~\rm{cm}^{-3}$ are shown.  Velocities are found using a linear fit of distance and time between 30--70 MHz}
\label{figure:bpl_v}
\end{figure}

Using the broken power-law, we ran simulations with varying initial beam densities to show the effect of changing $v_0$, the break energy.  Figure \ref{figure:bpl_v} shows how the velocities increase as $v_0$ is increased.  Injecting an electron beam with a high break energy has a significant effect on the resultant velocities of the electron beam.  This is not surprising given that we are significantly increasing the phase space density of the high-energy electrons by increasing $v_0$.  The simulation with $v_0=0.23$~c and $n_b=10^4~\rm{cm}^{-3}$ generated significant Langmuir waves with phase velocity $v_{\rm max}$.  Beam velocities would have been higher if our simulation box extended further than 0.7~c.

The effect of changing the break velocity, $v_0$, has been investigated by \citet{Li:2013aa}, showing that increasing the break velocity significantly increases both the peak velocity and the peak flux of fundamental radio wave emission.  We show here that $v_0$ will have a significant effect on the front and back velocities of an electron beam.  

The acceleration of a broken power-law is perhaps not a physically probable accelerated distribution.  There is no simple physical explanation why the accelerated electron distribution would form a plateau below $v_0$, unless caused by some other physical process like Langmuir wave generation.  A high break energy could be caused during particle acceleration in the presence of Coulomb collisions from a high temperature plasma.  Electrons below the break energy could be susceptible to a significant level of collisions with the background plasma within the acceleration timescale.  During the latter stages in solar flares, when the coronal plasma reaches temperatures in excess of 10~MK, collisions around the acceleration region will be significantly higher.

\subsection{Langmuir wave spectral energy density}

We calculated the front, peak and back of the electron beam from the derived fundamental emission brightness temperature (Equation \ref{eqn:t_b}) assuming a saturation of ion-sound waves.  The peak of the brightness temperature is proportional to $W_{\rm max}(k,r,t)/k^2$.  The rationale for using the radio brightness temperature is the desire to use type III bursts for remote sensing of electron beam properties.

Instead of using the radio brightness temperature we can instead use the Langmuir wave spectral energy density.  For a given time we find the value of $W_{\rm max}$ at each position, and then use the corresponding FWHM to find the front, peak and back of the electron beam.  The absence of $1/k^2$ means that Langmuir waves with higher $k$-vectors (lower phase velocities) play a more significant role in determining the front, peak and back of the beam.  

Another metric that can be used is the FWHM of the Langmuir wave electric field to find the front, peak and back of the electron beam.  The Langmuir wave electric field is a measurable quantity in the solar wind; unlike the Langmuir wave spectral energy density.  The electric field associated with the Langmuir waves can be found from $E^2(r,t)= 8\pi U_w(r,t)$ where $U_w(r,t)$ is the energy density of Langmuir waves, found by integrating $W(k,r,t)$ over $k$.  The electric field is even more influenced by Langmuir waves with higher $k$-vectors than the peak spectral energy density.

\begin{figure*}\center
\includegraphics[width=.9\textwidth,trim=10 25 20 0,clip]{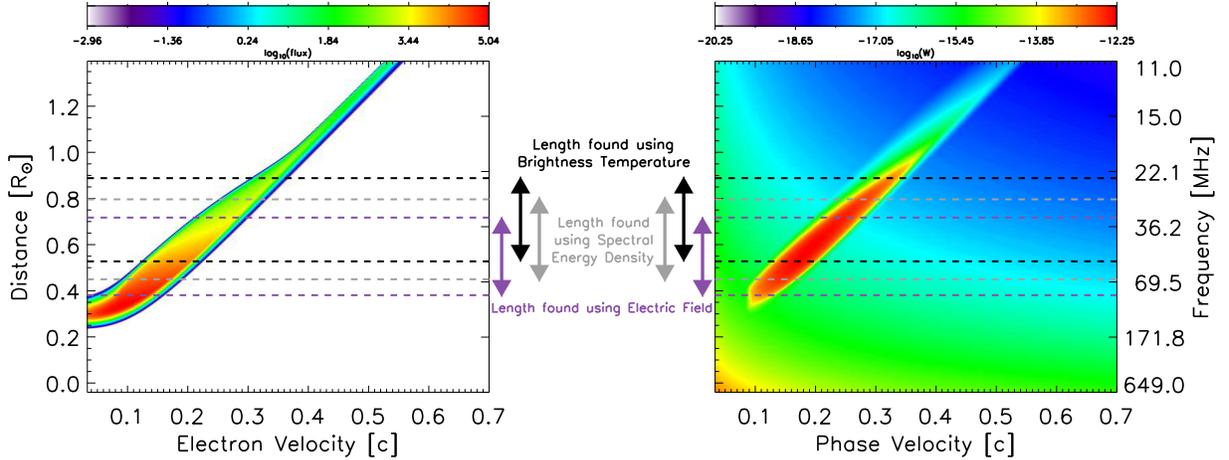}
\caption{The electron flux and Langmuir wave spectral energy density at 6 seconds for the initial electron beam parameters given in Table \ref{tab:beam_sun}.  The corresponding front and back of the electron beam is indicated by horizontal dashed lines, found from the brightness temperature (black), the peak value of the Langmuir wave spectral energy density (grey) and the Langmuir wave electric field (purple).}
\label{figure:eflux2}
\end{figure*}

We show the comparison between the front and back of the electron beam using the radio brightness temperature, the peak spectral energy density and the electric field in Figure \ref{figure:eflux2}.  The simulation is the same as shown in Figure \ref{figure:eflux} at $t=6$~seconds.  The difference between the three metrics is clear, with the method using radio brightness temperature being coincident with Langmuir waves at higher phase velocities and the method using the electric field being coincident with Langmuir waves at lower phase velocities.  The length of the electron beam using all three metrics remains similar, around $0.35~R_\odot$ at $t=6$~seconds.  The velocities attributed to the front and back of the electron beam are higher when the brightness temperature is used and lower when the electric field is used.

\subsection{Beam energy density and brightness temperature}

\begin{figure}\center
\includegraphics[width=0.49\textwidth,trim=20 0 0 0,clip]{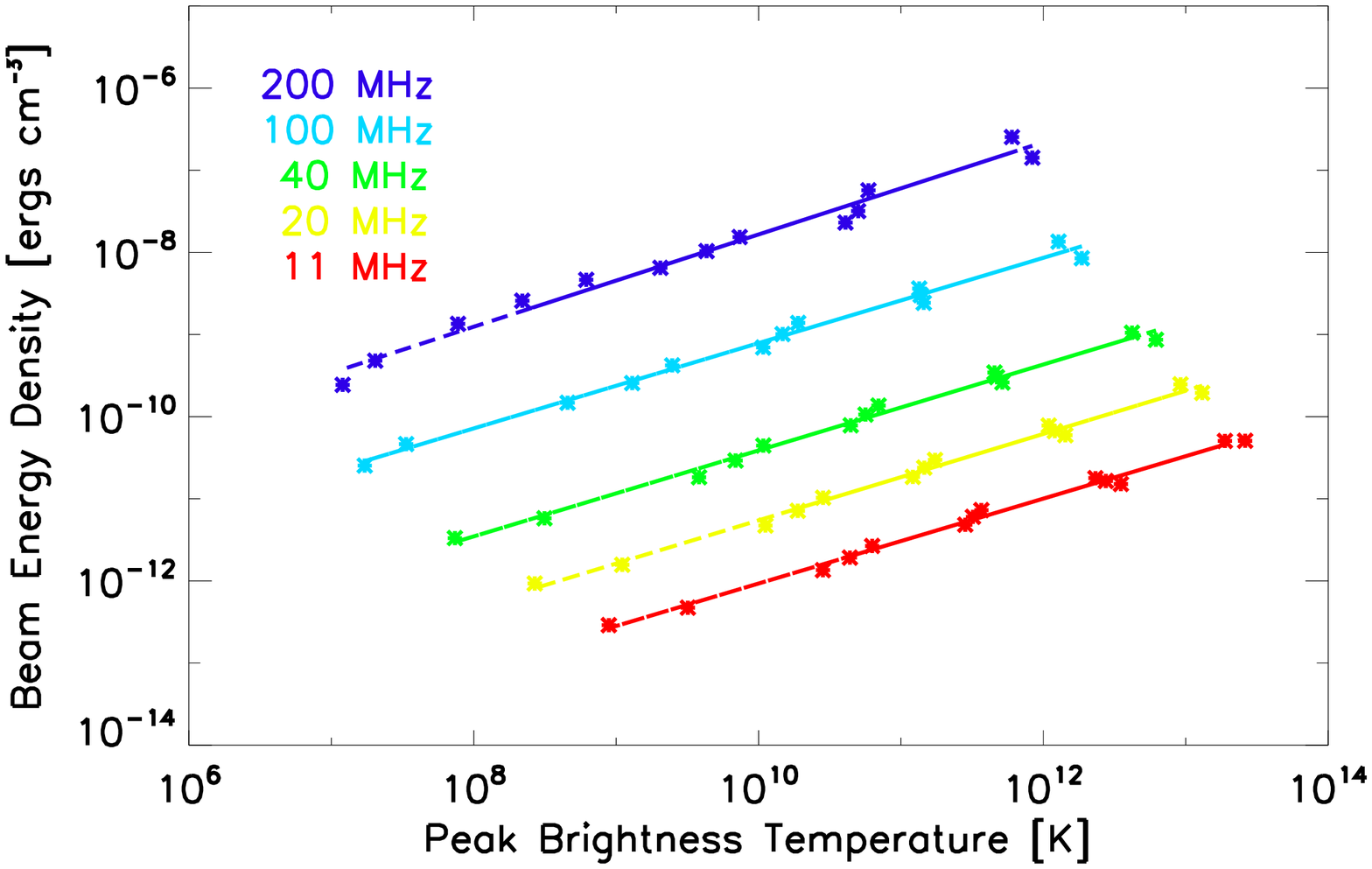}
\includegraphics[width=0.49\textwidth,trim=20 0 0 0,clip]{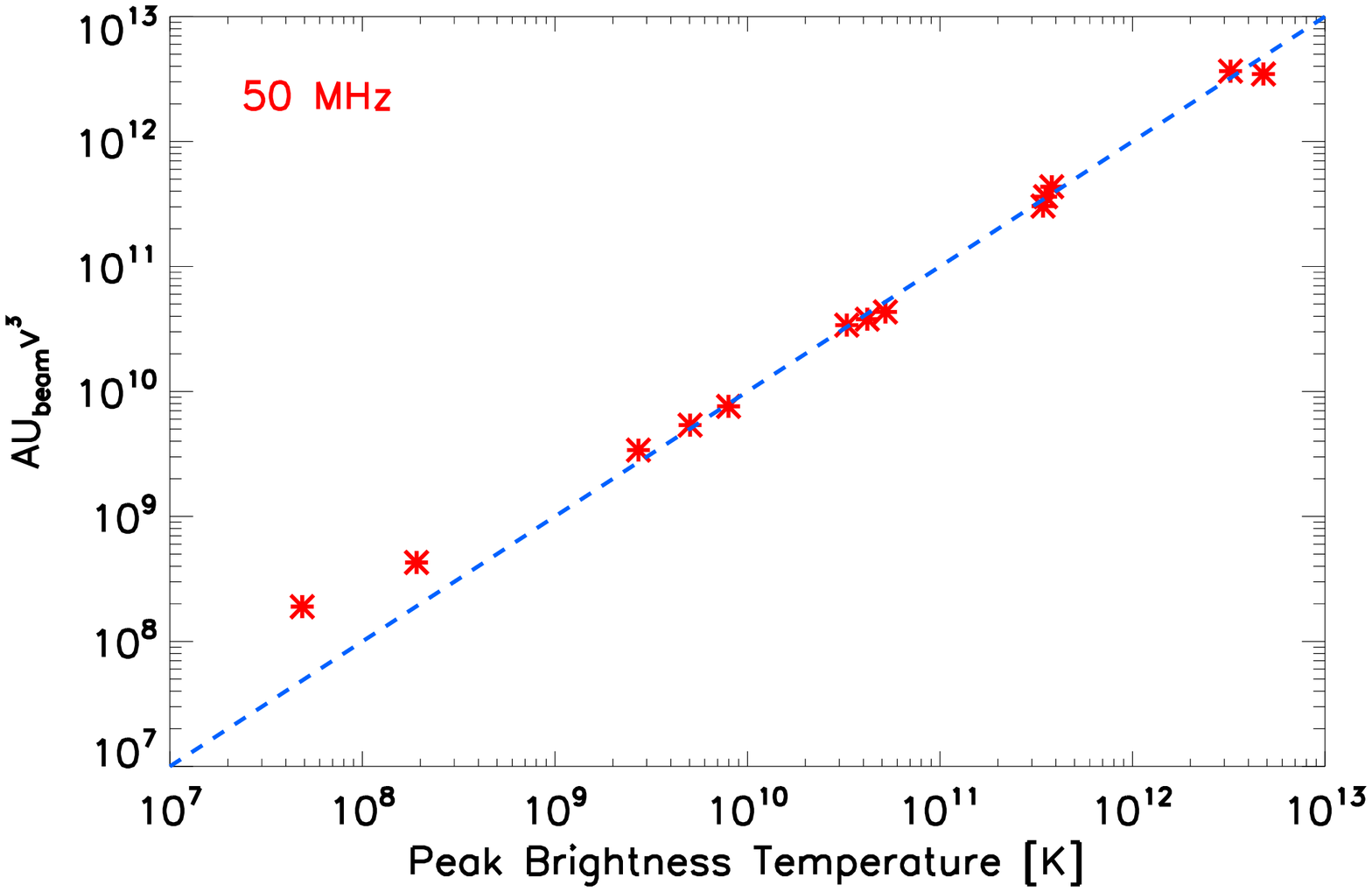}
\includegraphics[width=0.49\textwidth,trim=20 0 0 0,clip]{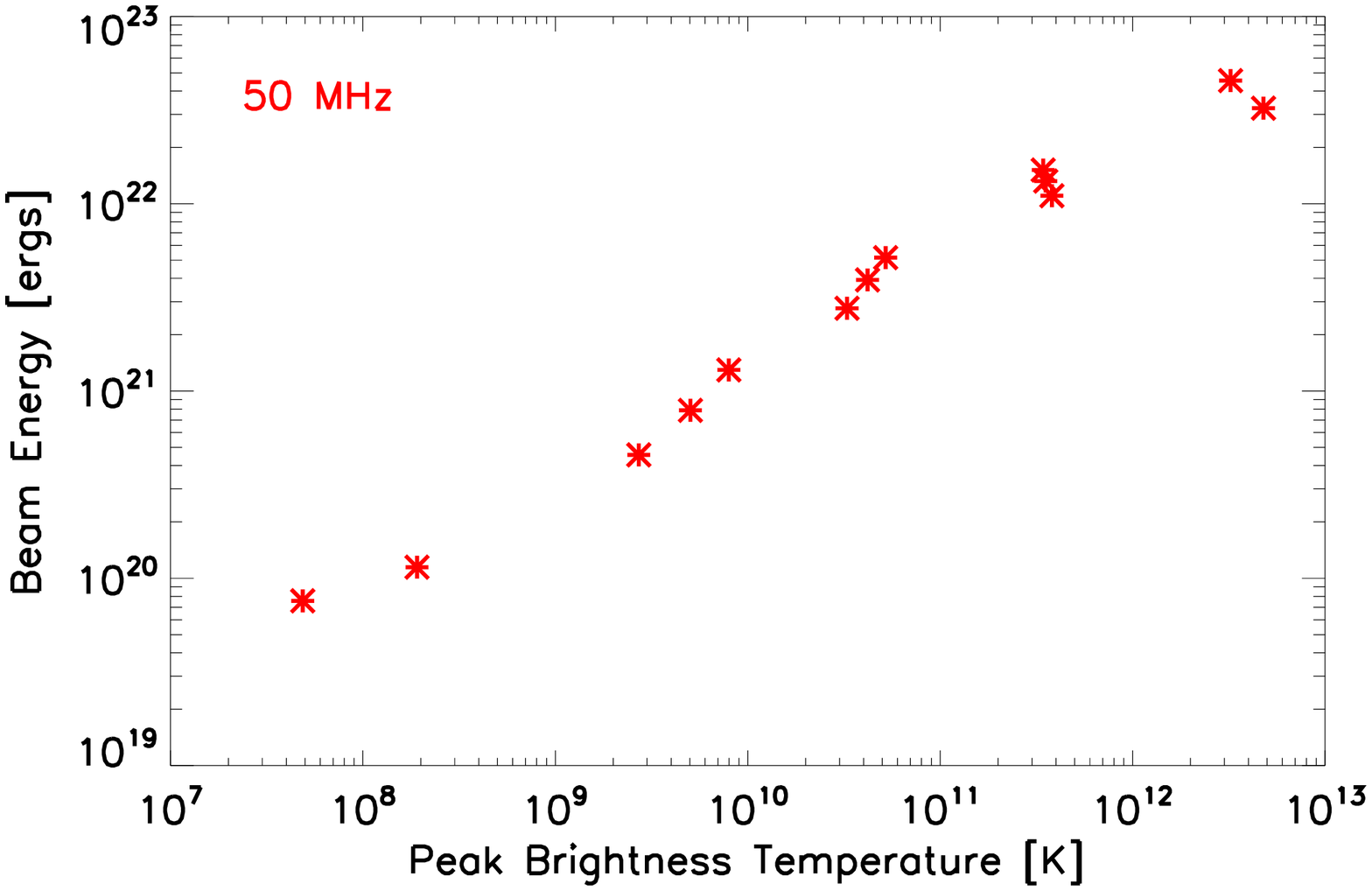}
\caption{Top: Peak beam energy densities as a function of peak brightness temperature for an initial electron beam with an instantaneous time injection.  Different points for a single frequency relate to beams with different initial beam densities and spectral indices.  Colours represent different frequencies.  Middle: Analytical estimates of the brightness temperature from gas-dynamic theory using the peak beam energy density $U_b$, the beam velocity at the peak brightness temperature and a constant of proportionality, $A$.  Bottom: Beam energy estimated from the energy densities at 50~MHz assuming a source size of 3.2 arcmins.  Note that comparisons of energy density and energy with observations can only be done for type IIIs with similar characteristics to our simulations.}
\label{figure:tb_bed}
\end{figure}

The complicated plasma emission mechanism has hampered using radio bursts as a diagnostic for electron beam energetics.  For the first time we present an estimate for the energy density contained within the electrons that produce the type III emission as a function of peak radio brightness temperature of a type III burst at different frequencies.  The electron energy density can be found from the electron distribution function $f(v,r,t)$ using
\begin{equation}
    U_{\rm beam}(r,t) = 0.5m_e\int_{v_{\rm  min}}^{v_{\rm  max}}{f(v,r,t)v^2dv}.
\end{equation}
For a given frequency, the peak energy density correlates to the peak brightness temperature, shown in Figure \ref{figure:tb_bed} using all simulations presented in Section \ref{sec:params}, which have different initial electron beam densities and velocity spectral indices.  At a single frequency, the greater the energy density of the electrons producing the Langmuir waves, the higher the radio brightness temperature they generate.  The energy density can be fit with a power-law at different frequencies, with the fit parameters given in Table \ref{tab:tb_bed}.  The spectral index at each frequency is approximately 0.5 but the amplitude decreases with decreasing frequency, $f$, and has the approximate relationship of $10^{-38.4}f^3$ for $f$ in Hz.  We can then approximate the energy density of the electrons responsible for the radio brightness temperature $T_T$ using
\begin{equation}\label{eqn:ub_tt}
    U_{\rm beam}(f,t) = 10^{-38.4}f^3T_T^{0.5}.
\end{equation}

% These two variables are related with a power-law with spectral index of approximately 0.5.  Table \ref{tab:tb_bed} contains the fit parameters for each power-law.  As the frequency decreases, the brightness temperature increases, as can be seen from Figure \ref{figure:dynspec}.  Similarly the beam energy density density decreases for decreasing frequencies, as the electron beam dilutes when it fills the radially expanding solar flux tube.  The decrease in the $\log$ amplitude with frequency, in Table \ref{tab:tb_bed}, has the approximate relationship of $3\log f-20.3$.

\vspace{20pt}
\begin{center}
\begin{table}
\centering
\caption{Fit Parameters used for beam energy density vs peak brightness temperature (Figure \ref{figure:tb_bed}).}
\begin{tabular}{ c  c  c  c  c  c}
\hline\hline
Frequency 	&  $\log$ Amplitude & Spectral Index \\ \hline
200 		& -13.4		& 0.56 	\\
100 		& -14.3		& 0.52 	\\
40 			& -15.6		& 0.52 	\\
20 			& -16.5		& 0.52 	\\
11 			& -17.2 	& 0.52 \\
\hline
\end{tabular}
\vspace{20pt}
\label{tab:tb_bed}
\end{table}
\end{center}

The correlation between beam energy density and brightness temperature can be explained using the analytical gas-dynamic solutions.  From \citet{Melnik:2000aa,Kontar:2001ac} we expect the peak Langmuir wave energy density to be
\begin{equation}
    W^{\rm max}(v) \propto m_e n_b v^3.
\end{equation}
We can substitute this into the beam brightness temperature, defined in Equation \ref{eqn:t_b}, obtaining 
\begin{equation} \label{eqn:Tb_v5}
    T_T^{\rm max}(r,t) \propto m_e n_b v^5.
\end{equation}
The relation in Equation \ref{eqn:Tb_v5} provides a close fit to the brightness temperature but an even closer fit is found using the beam energy density
\begin{equation} \label{eqn:Tb_v3}
    T_T^{\rm max}(r,t) \propto U_{\rm beam} v^3.
\end{equation}
The relation is shown in Figure \ref{figure:tb_bed} using a constant of proportionality, $A$.  The relation deviates slightly for the simulations with lower brightness temperatures.  The peak brightness temperature is thus heavily dependent upon the velocity (or energy) of the electrons that are generating the Langmuir waves; electron beams with higher velocities will typically have higher peak brightness temperatures.

Using the beam energy densities we can calculate the total energy of the electrons that are producing the type III emission, $E_{\rm beam}=U_{\rm beam}V$, by estimating the volume, $V$, of the beam.  For a given frequency, we can estimate the length of the beam in the direction of propagation using a sum of the duration, found from the FWHM of the brightness temperature, and the mean velocity of the electron beam, found from $\bar{v} = (v_{\rm front}+v_{\rm back})/2$.  We estimate the area of the beam using the cross-section of the magnetic flux tube that we use to model the $r^{-2}$ expansion of the electron beam travelling through the solar corona.  We assume an acceleration region cross-section of radius $d=10$~Mm which is at a distance of $30$~Mm along the cone of expansion.  The cross-section at 50~MHz, occurring at 413~Mm along the cone of expansion gives a radius of $10/30\times413=138$~Mm (3.2 arc minutes).  This is quite small, compared to observed source sizes \citep[e.g.][]{Kontar:2017ab}.  Faster expansion of the magnetic flux tube would increase source sizes but it would reduce the energy densities within the beam, at a given frequency.  

Assuming a Gaussian-distributed source we plot the energies obtained in Figure \ref{figure:tb_bed} as a function of peak brightness temperature.  As the energy contained in the deca-keV electrons increases, the corresponding radio brightness temperature increases.  The energies obtained are quite small, in comparison to those measured at 1~AU \citep{Krucker:2007aa,James:2017aa}.  However the durations of our electrons beams are significantly lower than those estimated at 1AU, with the number of electrons per second of our largest simulations being comparable to \citet{James:2017aa} at 74 keV.

The flux $S$ of our simulated radio bursts can be estimated using the source size $\theta$ of 3.2 arcmins at 50 MHz.  Using the Rayleigh-Jeans law 
\begin{equation}\label{eqn:tb_flux}
% S=2.65\left(\frac{\theta}{\rm{arc~mins}}\right)^2\left(\frac{\lambda}{\rm{cm}}\right)^{-2}\left(\frac{T_b}{\rm{K}}\right)
\left(\frac{T_b}{\rm{K}}\right) = 3770\left(\frac{S}{\rm{SFU}}\right)\left(\frac{\theta}{\rm{arc~mins}}\right)^{-2}\left(\frac{c}{\rm{cm~s^{-1}}}\right)^{2}\left(\frac{f}{\rm{Hz}}\right)^{-2},
\end{equation}
we find fluxes of 750 SFU for a brightness temperature of $10^{11}$~K.  However, this does not take into account any propagation effects like the scattering of light \citep[e.g.][]{Arzner:1999aa,Kontar:2017ab} that will occur from source to observer.  

We can construct an equation for the energy contained in the electrons responsible for the type III emission, based upon observable quantities.  The energy density of the beam can be found by combining Equation \ref{eqn:tb_flux} and Equation \ref{eqn:ub_tt}.  As described above, the volume of the beam can be found from the product of the length of the beam, $\bar{v}t_{\rm dur}$ and the source size, $\theta^2$.  Again, assuming that the intensity is Gaussian-distributed one arrives at the following equation 
\begin{equation}
\left(\frac{E_{\rm beam}}{\rm{ergs}}\right) =  8.8\times10^{-7}\left(\frac{\bar{v}}{\rm{cm~s^{-1}}}\right) \left(\frac{t_{\rm dur}}{\rm{s}}\right) \left(\frac{\theta}{\rm{arc~mins}}\right) \left(\frac{f}{\rm{Hz}}\right)^2 \left(\frac{S}{\rm{SFU}}\right)^{0.5} 
\end{equation}
where $E_{\rm beam}$ is the energy in the electrons responsible for the type III emission.  If applied to data, the above equation should only be used for small type III bursts that are similar to our simulations; type III bursts with very long durations of 10--100's seconds at 50~MHz are likely caused by long-duration injection profiles or multiple electron beams, that may change how the radio brightness temperature relates to beam energy density.

Our estimations serve as a first indication of the energy contained within an electron beam using type III bursts, where none currently exist.  However, one must be careful as our simulations do not take into account all of the physical processes that will affect the electrons during their propagation.  The beam energy estimates do not account for the turbulent background electron density \citep[e.g.][]{Melrose:1986aa,Kontar:2001ab,Reid:2010aa,Li:2012aa,Reid:2017ab}, that affects the level of Langmuir waves induced by an electron beam.  The simulations do not take into account the scattering of light from source to observer that can smear out the radio light curves in time \citep[e.g.][]{Arzner:1999aa,Kontar:2017ab}.  They do not model the pitch-angle scattering of electrons from magnetic fluctuations.

\subsection{Beam Duration}

\begin{figure}\center
\includegraphics[width=\wfig,trim=30 0 20 18,clip]{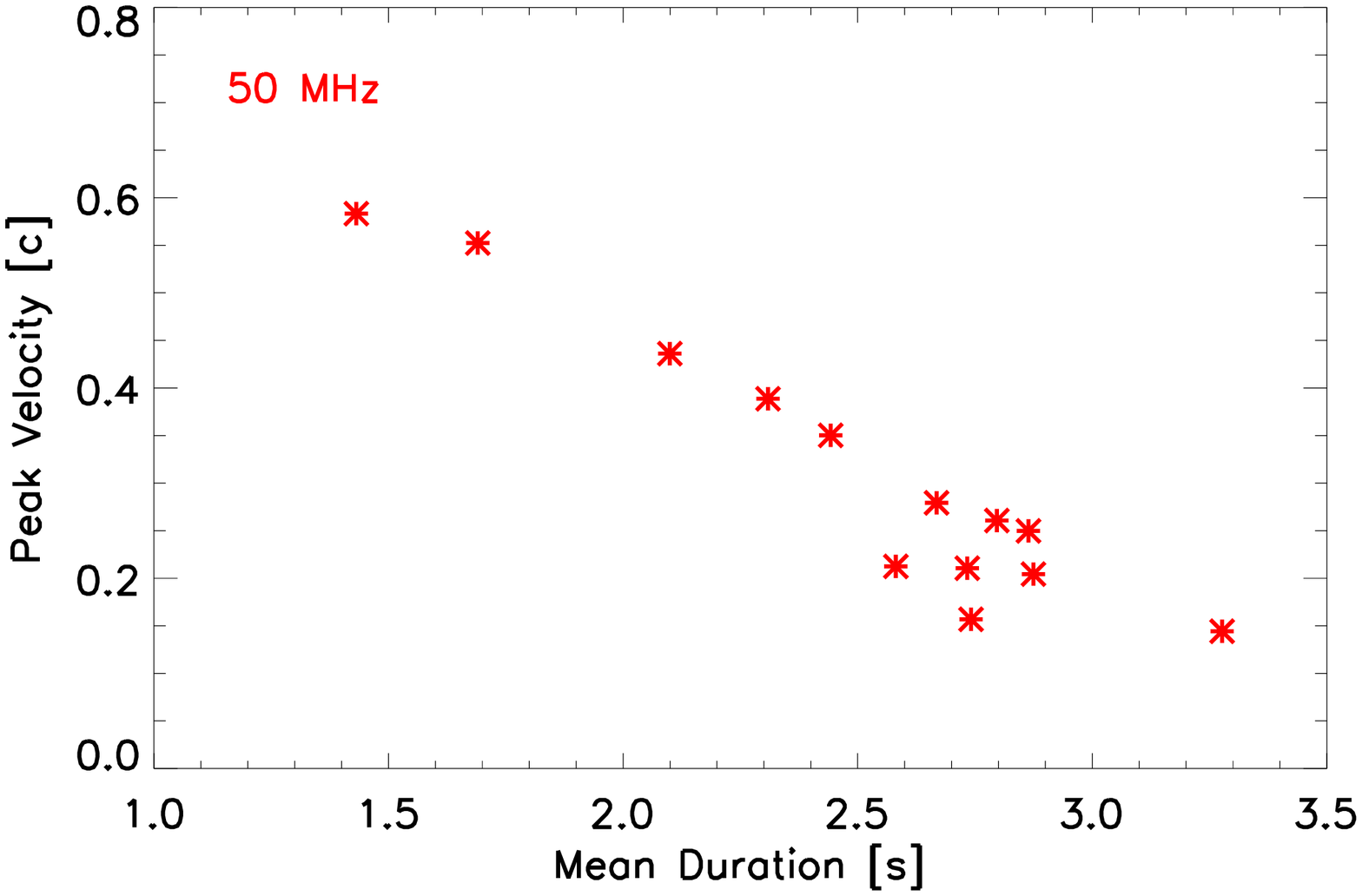}
\includegraphics[width=\wfig,trim=30 0 20 18,clip]{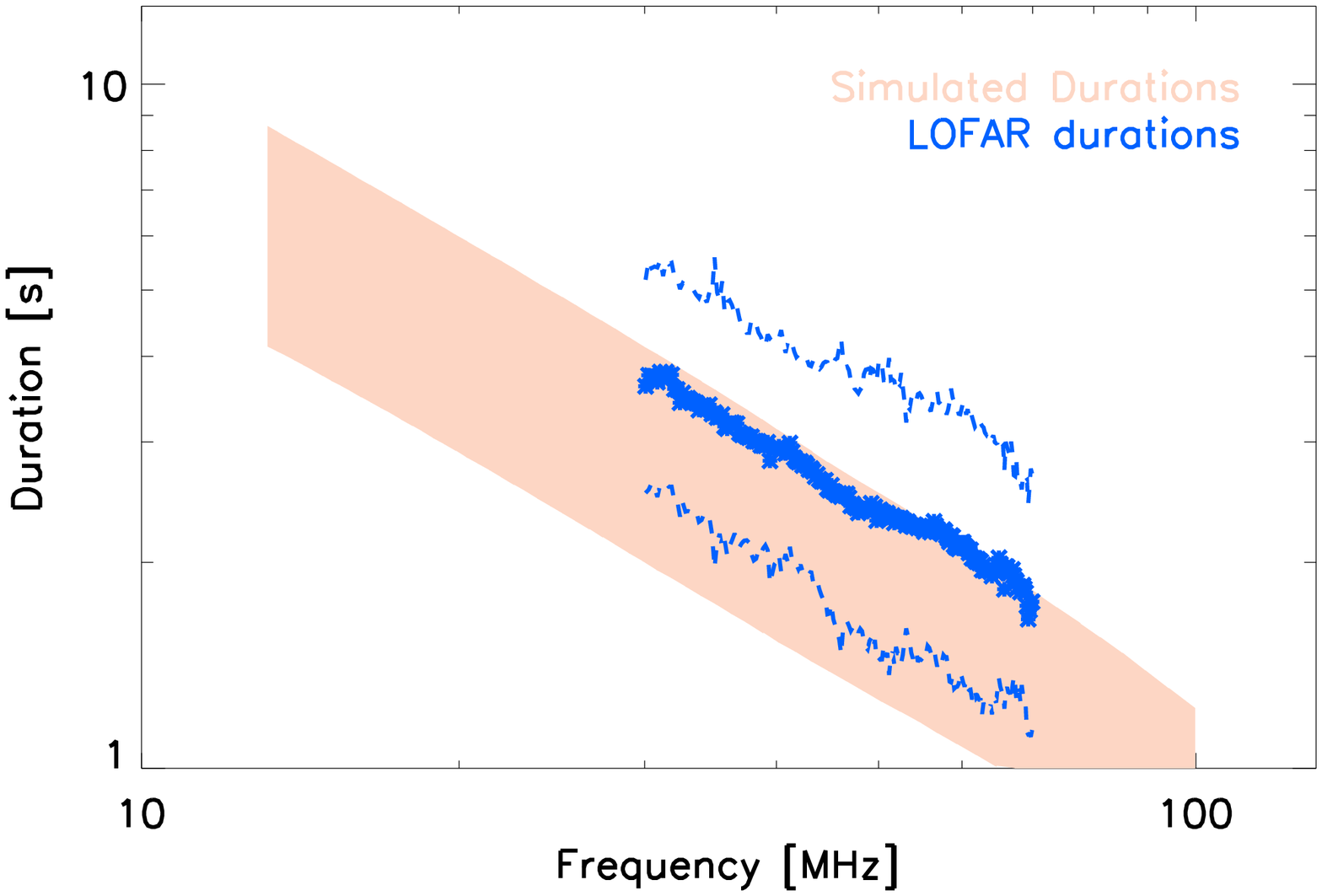}
\caption{Top: Mean type III duration as a function of velocity found from the motion of the peak brightness temperature.  Both variables were found between 30--70~MHz.  A strong correlation exists, in-line with observational results \citep{Reid:2018aa}.  Bottom: The range of duration as a function of frequency from the numerical simulations in the range 13--100 MHz.  Type III durations from \citet{Reid:2018aa} using LOFAR are overplotted, with the blue dashed lines showing the observed standard deviations.}
\label{figure:dur_vel}
\end{figure}

In a recent study, \citep{Reid:2018aa}, we analysed 31 type III bursts using LOFAR and quantified, among other variables, the FWHM duration of type III bursts lightcurves.  We found a strong correlation between the type III durations and the velocity of the exciters, derived assuming a density model and second harmonic emission.  We find the same correlation between the exciter velocities and the type III durations, using the simulated type III brightness temperatures, shown in Figure \ref{figure:dur_vel}.  The velocities found in  \citet{Reid:2018aa} from the radio bursts are lower than the peak velocities from most of our simulations, and we are finding the fundamental radio brightness temperature, but the correlation is still strong.  Our explanation in \citet{Reid:2018aa} holds, that the faster electrons, responsible for the faster derived velocities, have a shorter travel time through one point in space, and hence produce type III bursts with shorter durations.

The type III durations as a function of frequency were also shown in \citet{Reid:2018aa}.  For comparison we plot the range of durations found from the simulations as a function of frequency and overplot the type III durations observed using LOFAR.  There is a very good agreement in the durations.  The observed durations are at the higher end of the simulated durations, expected because the peak velocities estimated from the observations were at the lower end of the peak velocities found from the simulations.  The rise and decay times were also similar between the simulations and the observations, with the simulated rise times being smaller than the simulated decay times.

\section{Summary} \label{sec:conclusions}

We have simulated the propagation of electron beams through the solar corona, taking into account the resonant interaction with Langmuir waves in the background plasma.  We investigated how the front, peak and back of the electron beam evolve in time, as deduced from the peak type III brightness temperature.  We also showed how the electron beam energy density varies with brightness temperature of fundamental radio emission.

After injection of a power-law distribution in velocity, electrons initially propagate without generation of Langmuir waves. After the distance $x\simeq \delta d$, the first waves are generated. The velocity initially increases with distance as higher energy electrons produce significant Langmuir wave spectral energy density. As the injected distribution is a power-law, the maximum velocity increases as higher velocity electrons start to generate Langmuir waves. Later the velocity of the type III source decreases as a function of distance from the Sun.  The FWHM length of the source generating a type III burst increases as a function of time, with the rate of expansion (expansion velocity) being the difference between the front and back velocities.  The faster beams expand faster and the front of the electron beam moves faster than the peak which in turn moves faster than the back of the beam.

The initial electron distribution affects the type III properties, the velocities of type III bursts will be higher for higher initial electron beam densities and lower initial spectral indices (harder spectrum) of electrons.  Both parameters increase the energy density contained within the electrons that produce radio emission.  The higher the energy density in the beam, the  more energy was put into Langmuir waves and consequently the higher the derived brightness temperature.  

% The derived velocities were found to be influenced by the temperature of the background plasma or the minimum speed of the plateau formed during quasi-linear relaxation \citep{Kontar:1998aa,Li:2014aa}.  This is related to the idea that the background plasma stops Langmuir waves from being generated at lower phase velocities. The electron beam does not rise to high enough flux values above the background electron beam distribution.

Energetic electrons are also affected by the magnetic fluctuations that lead to the pitch-angle scattering of the electrons. This plays a role for the higher energy electrons that we are discounting when characterising the electron beams. There will also be some effect on the deca-keV electrons and its role on the electron transport has to be investigated. 

One of the important processes that affects the comparison of type III bursts and the electron simulation is the radio-wave propagation. Since the plasma emission originates at frequencies close to the plasma frequency, the radio waves will be strongly affected by scattering and refraction \citep[e.g.][]{Steinberg:1971aa,Riddle:1972aa,Arzner:1999aa}. Recent observations \citep{Kontar:2017ab} suggest that the spatial, and hence temporal characteristics of type III bursts at fundamental frequencies are influenced by radio-wave propagation effects. Specifically, the rise, the decay and the duration becomes dependent on the scattering of radio waves on the way from the radio source to the observer.  Therefore, the observed properties of type III bursts will be a convolution of electron transport studied in the paper and radio-wave transport. 

% \textbf{There should be a discussion on the generation of second harmonic radio emission from the Langmuir waves.  This should highlight the differences in the front, peak and back of the electron beam, as found using second harmonic emission.  Crucially whether lower of higher velocities will be derived compared to the fundamental emission.}

% \textbf{This section should discuss the different types of information that you can discern from the plasma emission.  One of the main parts is that it should be potentially possible to create a way in which you can predict the beam parameters from the dynamic spectrum of a type III burst.  At least a look-up table would be a real improvement over what is currently available.  Even if this is an approximation, it will help use radio in conjunction with other wavelengths.}

With the impending launches of Parker Solar Probe and Solar Orbiter we will be able to measure in situ the electron beams close to the Sun as they produce radio emission.  Whilst they will not observe in situ at frequencies above 10 MHz, the combined observations of energetic electrons, Langmuir waves and Type III bursts should be able to test the velocities at different parts of the electron beam, and estimate beam length.  A great many more electron beams are likely to be observed close to the Sun than at 1~AU, in particular with accompanying Langmuir waves, as beam densities will not have decreased as much in the expanding solar wind plasma.  Related to this fact, we predict that higher energy electrons ($>10$~keV) will be measured co-temporal with Langmuir waves closer to the Sun, especially at the front of the electron beam.  Moreover the beams that produce type III radio emission are likely to have higher fluxes than beams where radio emission is absent, for a given energy.  Finally, Solar Orbiter should detect the transport change in the beam spectral index by combining the X-ray measurements with in situ data. 

% \textbf{Additionally, there could be a section highlighting the predictions for Solar Orbiter and Parker Solar Probe.  Specifically, that the electrons which arrive co-temporally with the increase in the electric field due to Langmuir waves will be at higher velocities closer to the Sun.  For type III events, when Langmuir waves are observed at the same time as electrons in one specific energy channel, the flux of electrons will be higher than when Langmuir waves are no observed.}

\begin{acknowledgements}
We acknowledge support from the STFC consolidated grant ST/P000533/1. This work benefited from the Royal Society grant RG130642. H. Reid acknowledges Paulo Simoes for useful discussions on the presentation of the results.
\end{acknowledgements}

\bibliographystyle{aasjournal}
% \bibliography{/Users/hamish/Documents/papers/ubib}
\bibliography{ubib}

\end{document}